\documentclass[aps,reprint,superscriptaddress,amsmath]{revtex4-1}

\pdfoutput=1

\usepackage{upgreek,graphicx}

\newcommand{\Dss}{\Delta_\textrm{ss}}
\newcommand{\Dps}{\Delta_\textrm{ps}}

\begin{document}

%TITLE
\title{Holey Random Walks: Optics of Heterogeneous Turbid Composites}

% AUTHORS
\author{Tomas Svensson}
\email[Email: ]{svensson@lens.unifi.it}
\affiliation{European Laboratory for Non-linear Spectroscopy (LENS), University of Florence, Via Nello Carrara 1, 50019 Sesto Fiorentino (FI), Italy}

\author{Kevin Vynck}
\affiliation{European Laboratory for Non-linear Spectroscopy (LENS), University of Florence, Via Nello Carrara 1, 50019 Sesto Fiorentino (FI), Italy}

\author{Marco Grisi}
\affiliation{European Laboratory for Non-linear Spectroscopy (LENS), University of Florence, Via Nello Carrara 1, 50019 Sesto Fiorentino (FI), Italy}

\author{Romolo Savo}
\affiliation{European Laboratory for Non-linear Spectroscopy (LENS), University of Florence, Via Nello Carrara 1, 50019 Sesto Fiorentino (FI), Italy}

\author{Matteo Burresi}
\affiliation{European Laboratory for Non-linear Spectroscopy (LENS), University of Florence, Via Nello Carrara 1, 50019 Sesto Fiorentino (FI), Italy}
\affiliation{Istituto Nazionale di Ottica (CNR-INO), Largo Fermi 6, 50125 Firenze (FI), Italy}

\author{Diederik S. Wiersma}
\affiliation{European Laboratory for Non-linear Spectroscopy (LENS), University of Florence, Via Nello Carrara 1, 50019 Sesto Fiorentino (FI), Italy}
\affiliation{Istituto Nazionale di Ottica (CNR-INO), Largo Fermi 6, 50125 Firenze (FI), Italy}

% DATE
\date{\today}

% ABSTRACT
\begin{abstract}
We present a probabilistic theory of random walks in turbid media with non-scattering regions. It is shown that important characteristics such as diffusion constants, average step lengths, crossing statistics and void spacings can be analytically predicted. The theory is validated using Monte Carlo simulations of light transport in heterogeneous systems in the form of random sphere packings, and good agreement is found. The role of step correlations is discussed, and differences between unbounded and bounded systems are investigated. Our results are relevant to the optics of heterogeneous systems in general, and represent an important step forward in the understanding of media with strong (fractal) heterogeneity in particular. 
\end{abstract}

% PACS numbers (not shown in arXiv version)
% 42.25.Dd       Wave propagation in random media
% 05.40.Fb 	Random walks and Levy flights 
% 42.68.Ay 	Propagation, transmission, attenuation, and radiative transfer (see also 92.60.Ta Electromagnetic wave propagation)
% 05.10.Ln 	Monte Carlo methods 
% 05.40.Jc 	Brownian motion

\maketitle

% INTRODUCTION
\section{Introduction}

Multiple scattering of light is ubiquitous in nature, and a proper account of the phenomenon is essential in important areas such as atmospheric science \cite{Thomas2002_Book,Mishchenko2006_Book,Davis2010_RepProgPhys}, biomedical optics \cite{Tuchin2007_Book,Welch2010_Book,Wang2007_Book} and material characterization \cite{Berne2000_Book,Reich2005_AdvDrugDeliverRev,Siesler2008_Book,Shi2010a_JPharmSci}. Despite the abundance of situations in which scatterers are not uniformly distributed, multiple scattering of light is often treated within the framework of radiative transfer and diffusion theory while assuming exponentially distributed ballistic segments. At the same time, significant efforts have also been directed towards the understanding of heterogeneous systems and non-Poissonian transport mechanisms \cite{Shaw2002_JQSRT,Davis2004_JQSRT,Davis2011_JQSRT,Bal2011_JQSRT}. In fact, the presence of heterogeneities has important implications in a wide variety of contexts, from the optics of cloud systems \cite{Lovejoy1990_JGeophysRes,Davis1999_ARM_Proc,Scholl2006_JGeophysRes,Davis2010_RepProgPhys} to spectroscopy of the human body \cite{Liu1995_MedPhys,Firbank1996_PhysMedBiol,Hielscher1998_PhysMedBiol,Boas2002_OptExpress,Bal2002_JCompPhys,Shah2004_JBO,Gibson2005_PhysMedBiol,Ntziachristos2005_NatBiotechnol,Svensson2007_JBO}, to transport in colloids and foams \cite{Dogariu1992_WavesRandomMedia,Sorensen2001_AerosolSciTech,Durian1991_Science,Gittings2004_EPL}. However, since light transport in such cases is intrinsically related to the specific heterogeneity, its description in general terms remains elusive.

Here, we present a probabilistic theory for light transport in turbid media with non-scattering regions inside. We show that important characteristics such as asymptotic diffusion constants, step distributions and void crossing statistics can be analytically predicted via simple formulas. The success of the theoretical approach is verified by analysis of random walks in random three-dimensional polydispersive sphere packings (the inter-sphere volume set to be homogeneously turbid). Our findings provide important insight on light transport in materials with spatial heterogeneity in scatterer density, including complex porous materials and disordered optical materials with engineered fractal heterogeneity such as L\'evy glasses \cite{Barthelemy2008_Nature,Burioni2010b_PRE,Barthelemy2010_PRE,Buonsante2011_PRE,Burresi2012_PRL,Groth2012_PRE}.

% HOLEY RANDOM WALKS
\section{Analytical description of holey random walks}

\begin{figure}[h]
  \includegraphics[]{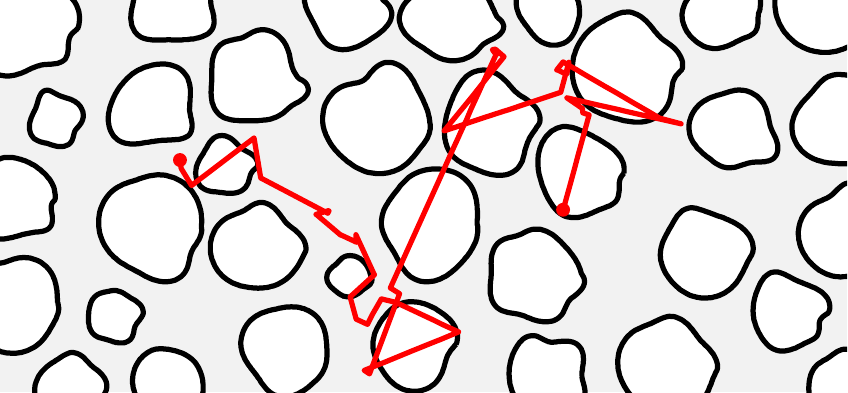}\\
  \caption{A holey random walk. Random walkers change direction randomly in time, but only when traveling in the turbid medium found in between the non-scattering regions (holes). Transport is governed by the resulting step length distribution and step correlations in a complex manner.}\label{FIG1_HoleyRandomWalk}
\end{figure}

Let us start to consider the general case of a random walk in a composite media consisting of arbitrarily shaped non-scattering regions embedded in a turbid medium (a holey system), the void filling fraction being $\phi$ (cf. Fig. \ref{FIG1_HoleyRandomWalk}). 

We will assume that isotropic scatterers are randomly distributed in the turbid medium, meaning that the distance between isotropic scattering events is exponentially distributed with a scattering mean free path $\ell_s$. The scattering mean free path is related to single scattering properties via 
\begin{align}\label{eq:l_s}
    \ell_s=\frac{1}{n\sigma_s}
\end{align}
where $n$ denotes the number density of scatterers and $\sigma_s$ the scattering cross section. In general, the random step length $S$ between two subsequent scattering events in the composite heterogeneous medium consists of two parts: a length $S_\textrm{turbid}\in\textrm{Exp}(\ell_s)$ inside the turbid inter-void medium, and a length $S_\textrm{void}$ inside non-scattering void(s):
\begin{align}\label{eq:S}
    S = S_\textrm{turbid} + S_\textrm{void}.
\end{align}
Here, it should be noted that $S_\textrm{turbid}$ and $S_\textrm{void}$ are not independent random variables. On the contrary, they are positively correlated: a long step in the turbid medium is more likely to be accompanied by one or several crossings of non-scattering regions.

%EQUILIBRIUM CONSIDERATIONS
\subsection{Equilibrium considerations}

Assuming that the density of states is homogenous (which in optics corresponds to a constant refractive index), random walkers at equilibrium should sample the two constituents according to their respective volume fraction. We then expect that a fraction $\phi$ of the average step is located to voids, i.e.
\begin{align}%\label{eq:Equilibrium_relations}
    & E[S_\textrm{void}]   = \phi E[S]  \label{eq:Equilibrium_relation_void} \\
    & E[S_\textrm{turbid}]  = (1-\phi) E[S] \label{eq:Equilibrium_relation_turbid}
\end{align}
In terms of expectations values, we have
\begin{align}\label{eq:E_S}
   E[S] = E[S_\textrm{void}] + E[S_\textrm{turbid}],
\end{align}
and using that we per definition also know that $E[S_\textrm{turbid}] =\ell_s$, we reach $E[S] = \phi E[S] + \ell_s$ and conclude that
\begin{align}\label{eq:E_S_equals_l_h}
    E[S] = \frac{\ell_s}{1-\phi} = \ell_h
\end{align}
This important result that tells us two things: the mean step length is independent of the shape and size distribution of the non-scattering regions, and it equals the homogenized scattering mean free path $\ell_h$ that would govern a system where the scatterers were randomly distributed over the full volume instead of only a volume fraction $1-\phi$ ($n$ going from $n_0$ to $n_0(1-\phi)$ in  Eq. \ref{eq:l_s}). $E[S]$ is thus dependent only on $\phi$, not on the sizes of the voids. This means that the mean step will only grow with an increase of $\phi$, and not with the inclusion of larger and larger voids.

\subsection{A general remark regarding diffusion}

Although the mean step is a very important quantity, transport properties are determined by the step length distribution as a whole. In particular, when steps can be considered independent, the diffusion constant that governs the macroscopic spreading of random walkers is determined by the ratio of the first two moments of the step distribution (when finite). This important relation can easily be derived by considering the summation of independent $d$-dimensional isotropic random steps, all following the same (arbitrary) step length distribution. One finds that the mean square displacement asymptotically grows as $2dDt$ with a diffusion constant $D$ given by
\begin{align}\label{eq:Diffusion_constant}
    D = \frac{1}{2d} \times v \times \frac{E[S^2]}{E[S]},
\end{align}
where $S$ denotes the scalar length (norm) of the individual steps, and $v$ is the walk velocity. For purely exponential steps in 3D, corresponding to a homogenous distribution of scatterers, this expression reduces into the famous relation
\begin{align}\label{eq:Diffusion_constant_exp}
    D = \frac{1}{3} v \ell_t
\end{align}
where $\ell_t$ is the so called transport mean free path (the mean length of the exponential steps). It should be noted that this relation is normally derived via the radiative transfer equation \cite{Lagendijk1996_PhysRep,Rossum1999_RevModPhys}, not by considering summation of random variables (but see, e.g., \cite{Bouchaud1990_PhysRep,Ben-Avraham2000} for similar considerations). For a general holey system, with heterogeneous distribution of scatterers, steps are not exponential distributed and the simple formula of Eq. \ref{eq:Diffusion_constant_exp} does not hold. Interestingly, we have already shown that the mean step length, $E[S]$, of the holey system is independent of the scatterer distribution. The second moment, $E[S^2]$, will on the other hand depend on the particular heterogeneity and scatterer density in a complex manner.

% SPHERE PACKINGS AS HOLEY SYSTEMS
\subsection{Sphere packings as holey systems}

As indicated above, transport in a general holey system is complicated. Beside the issue of step length distribution, step correlations may be important and must not be forgotten. To gain insight on this complicated matter we will turn to systems where the non-scattering part of the system have the form of a polydispersive random sphere packing. In this and the following section, we will present an analytical theory of transport in such systems. In subsequent sections, this theory will be compared to Monte Carlo simulations of random walks in sphere packing realizations.

\begin{figure}[h]
  \includegraphics[]{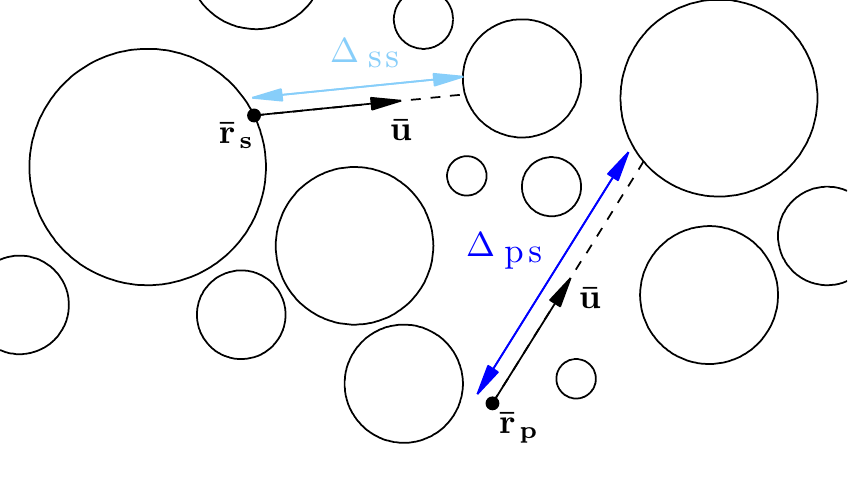}\\
  \caption{From a random walk perspective, two central characteristics of the sphere packing are the distribution of (i) the distance from a random point $\mathbf{\bar{r}_p}$ (scattering event) along a random direction $\mathbf{\bar{u}}$ to a sphere surface, $\Dps(\mathbf{\bar{r}_p},\mathbf{\bar{u}})$, and (ii) distance to the next sphere along a random direction when leaving a sphere at a random point $\mathbf{\bar{r}_s}$, $\Dss(\mathbf{\bar{r}_s},\mathbf{\bar{u}})$. Their distributions can be assessed via statistical analysis of sphere packings. Later, we show that the mean value of $\Dss$, as experienced by the random walker, can be analytically predicted.}\label{FIG2_Dss_Dps_Illustration}
\end{figure}

Let $r_i$ ($i=1,\ldots,M$) denote the different radii of the involved spheres, and $n_i$ their respective number density. A random walker traveling in this system will scatter randomly in time, but only when being outside of the non-scattering regions. When being scattered, the chance of crossing at least one sphere in the coming step depends on the probability that the next step is longer than the distance to the next sphere surface (along the new walker direction). This distance can be seen as a random variable, and is here denoted $\Dps$ (ps as in point-sphere). If a sphere is crossed, the probability to cross also a second sphere before being scattered now depends on the sphere-sphere spacing along the direction of the walker. Also this distance is a random variable, and we denote it $\Dss$ (ss as in sphere-sphere). The actual step length distribution will be determined by the distributions of these random variables (of course, in combination with the inter-sphere scattering mean free path $\ell_s$). Fig. \ref{FIG2_Dss_Dps_Illustration} illustrates the definition of these two essential distance variables.

The average spacing between spheres $\Dss$, as experienced by the random walker, can be analytically calculated from the filling fraction and the sphere distribution. If we consider a random line drawn through the sphere packing, it is evident that a fraction $1-\phi$ of it will be drawn through the intersphere medium. Letting $E[\zeta]$ denote the average length of the individual chords through the spheres, we have that
\begin{align}\label{eq:Mean_Dss}
    \frac{E[\Dss]}{E[\Dss]+E[\zeta]} = 1-\phi 
\end{align}
and, similarly to Olson et al. \cite{Olson2006_JQSRT}, conclude that
\begin{align}\label{eq:Mean_Dss_2}
    E[\Dss]=\frac{1-\phi}{\phi}E[\zeta].
\end{align}
The average chord length $E[\zeta]$ can, in turn, be calculated from the sphere distribution. Let $p_i$ denote the probability that a random chord is made in a sphere of radius $r_i$. Intuitively, this probability should be proportional to the total surface area of the sphere category, i.e.
\begin{align}\label{eq:p_i}
    p_i=\frac{A_i}{\sum\limits_{j} A_j} = \frac{n_i\times r_i^2}{\sum\limits_{j} n_j\times r_j^2}
\end{align}
In fact, this view is equivalent to the equilibrium argument adhered to in this article: the total path made through the different sphere categories should equal their fraction of the total volume. Assuming an isotropic flux of random walkers close to the sphere surfaces (see discussion in \cite{Kruijf2003_AnnNuclearrEnerg}), the length of a chord through a sphere of radius $r_i$ follows a triangular probability density function $f_i(x) = x/(2r_i^2)$ in $0\leq x \leq  2r_i$. The average chord being $\ell_i=4r_i/3$ and the mean square chord being $2r_i^2$. The probability density function of a random chord $\zeta$ in the polydispersive packing, $f_\zeta(x)$, is a weighted sum of the individual chord distributions $f_i$:
\begin{align}\label{eq:pdf_chord}
    f_\zeta(x) = \sum\limits_i p_i \times f_i(x) = \sum\limits_{i:2r_i>x} p_i \times \frac{x}{2r_i^2}.
\end{align}
The mean overall chord then becomes
\begin{align}
E[\zeta]=\sum p_i\ell_i = \sum p_i\times\frac{4r_i}{3},
\end{align}
and its mean square
\begin{align}
E[\zeta^2]=\sum p_i \times 2r_i^2.
\end{align}

% THE EXPONENTIAL SPACING MODEL
\subsection{The exponential spacing model}\label{sec:exp_spacing_model}

From statistical analysis of random sphere packings, we generally find that $E[\Dps]\approx E[\Dss]$ and that distributions have near-exponential decay. We therefore propose a model in which the actual (complicated) distributions of $\Dps$ and $\Dss$ are modeled with a \textit{single} exponentially distributed random spacing $\Delta$ with mean spacing $\bar{\Delta}$, i.e. $\Delta\in\textrm{Exp}(\bar{\Delta})$. In fact, it has been shown that spacing between random objects is, to a very good approximation, exponential at low filling fractions \cite{Olson2006_JQSRT}. At higher filling fractions, spacing distributions are no longer be perfectly exponential, but rather a complicated function of the exact structure \cite{Torquato1993_PRE,Gueron2003_PRE,Olson2006_JQSRT}. But also in such cases, we will show that the errors due to the use of a simple exponential spacing model can be reasonably small. 

Within the exponential spacing model, the probability that an exponentially distributed step $S_\textrm{turbid}\in\textrm{Exp}(\l_s)$ will not take the random walker to a sphere is 
\begin{align}\label{eq:P_0}
   P_0=P(S_\textrm{turbid}<\Delta)=\frac{\bar{\Delta}}{\bar{\Delta}+\ell_s}.
\end{align}
Given the memoryless property of the Poisson process that governs scattering events, the number of crossings naturally follows a geometric distribution. This means that the probability to cross $n$ spheres in between two scattering events is
\begin{align}\label{eq:P_n}
   P_n = (1-P_0)^nP_0.
\end{align}
The average number of sphere crossings per step $N$ is, within this model,
\begin{align}\label{eq:N_cross_mean_exp_model}
   E[N]=\sum n\times P_n=\frac{1-P_0}{P_0}=\frac{\ell_s}{\bar{\Delta}}
\end{align}
In order to fulfill the equilibrium condition, $\bar{\Delta}$ must be chosen so that the average step samples the medium according to volume fractions. Assuming that the number of sphere crossings and involved chord lengths are independent, we should have that
\begin{align}\label{eq:PathFractionl}
   \frac{E[N]E[\zeta]}{E[N]E[\zeta]+\ell_s}=\phi
\end{align}
Using Eq. \ref{eq:N_cross_mean_exp_model}, we then reach the conclusion that 
\begin{align}\label{eq:MeanSpacing}
   \bar{\Delta} = E[\Dss]=\frac{1-\phi}{\phi}E[\zeta]  
\end{align}

The exponential spacing model we have now outlined allows us to the estimate step length distribution. In particular, we can use it to estimate the mean squared step $E[S^2]$. Again, we will assume that, for each step, the involved chord lengths are independent of the number of sphere crossings. By splitting the outcome of $S^2$ into the number of crossings and summing the conditional expectations (details in Appendix \ref{app:E_SSq}), we find that
\begin{align}\label{eq:SSq_summary}
   E[S^2] &= E[S^2_\textrm{turbid}] + E[S^2_\textrm{void}] + 2E[S_\textrm{turbid}S_\textrm{void}]  \nonumber \\
    &= 2\ell_s^2 \nonumber \\
    & \quad + \sum\limits_{n=0}^{\infty} P_n \times (nE[\zeta^2] + n(n-1)E[\zeta]^2) \nonumber \\ 
    & \quad + 2\sum\limits_{i=0}^\infty P_n\times nE[\zeta] \times(n+1) \frac{\bar{\Delta}\ell_s}{\bar{\Delta}+\ell_s} \nonumber \\
    &= 2\ell_h^2 + \phi\ell_h\times \frac{E[\zeta^2]}{E[\zeta]}
\end{align}

% THE DIFFUSION CONSTANT
\subsection{The diffusion constant for holey system}

Assuming that step correlations are negligible, and remembering that $E[S]=\ell_h$, the expression for $E[S^2]$ given above (Eq. \ref{eq:SSq_summary}) allows us to write a closed form expression for the diffusion constant that governs macroscopic transport. Introducing the homogenized diffusion constant
\begin{align}\label{eq:D_h}
    D_h = \frac{1}{3}v \ell_h
\end{align}
and the "chord-domain" diffusion constant (the diffusion constant for a random walk with steps purely following the chord step distribution)
\begin{align}\label{eq:D_chord}
    D_\zeta = \frac{1}{6}v \times \frac{E[\zeta^2]}{E[\zeta]}
\end{align}
we reach, via Eq. \ref{eq:Diffusion_constant}, the surprisingly simple relation
\begin{align}\label{eq:D_holey}
   D = D_h + \phi D_{\zeta}
\end{align}
Interestingly, the diffusion constant of the heterogeneous system is given by the homogenized diffusion constant plus a term which is independent of the inter-void scattering mean free path $\ell_s$. Note also that, as expected, the expression reduces to the homogenized diffusion constant as $\phi$ approaches zero.

% MONTE CARLO SIMULATIONS
\section{Monte Carlo simulations of transport in quenched disorder}

\begin{figure}[ht]
  \includegraphics[]{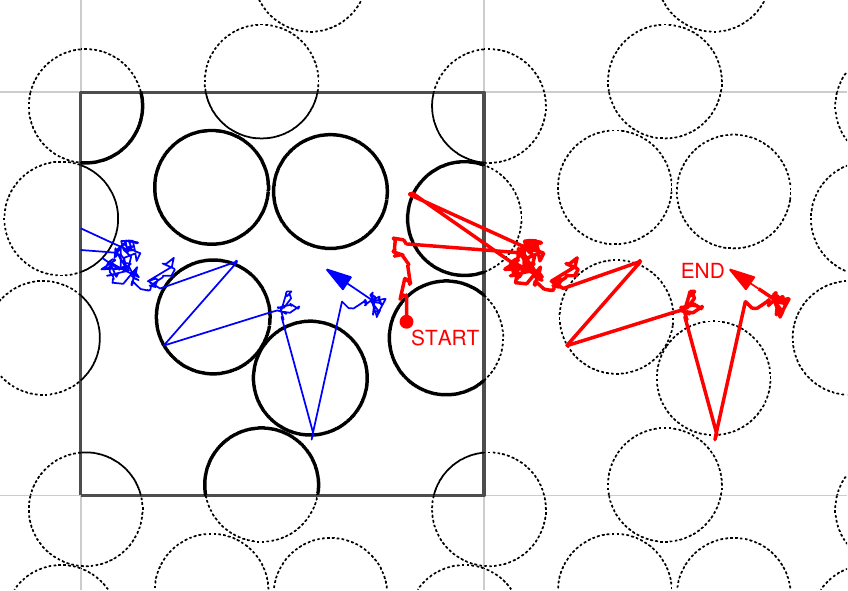}\\
  \caption{2D illustration of unbounded holey random walks based on a sphere packings with periodic boundary conditions. When a walker leaves the sphere packing unit cell (marked square), the walk continues on the opposite side of the same sphere packing. The path of interest (red) is reconstructed by keeping track of boundary crossings, but path outside the unit cell is in reality a path inside the one and only unit cell (blue path). The use of periodic boundary conditions in this manner enables the use of realistically sized sphere packings, efficient averaging over disorder realization by allowing walkers to start also close to the sphere packing boundaries, and long-time tracking of spreading. At the same time, the size of the unit cell is made large enough to make effects of periodicity negligible.}\label{FIG3_PeriodicRandomWalk}
\end{figure}

To illustrate the validity and importance of the theory presented in earlier sections, this section will compare it with Monte Carlo (MC) simulations of transport in quenched random sphere packings. We use the term \textit{quenched} to emphasize that the random walk is done in the quenched (frozen) heterogeneity of a random sphere packing, and not in fully annealed disorder model. The sphere packings used in our simulations are created by random sequential addition of spheres \cite{Torquato2001_Book} in descending order of size. Moreover, they are created to have periodic boundary conditions, i.e. so that the complete (cubic) sphere packing can act as a unit cell that can be stacked in all directions. As illustrated in Fig. \ref{FIG3_PeriodicRandomWalk}, random walks are performed so that when crossing a boundary of the unit cell, the random walk can be continue on the opposite side of the same unit cell (while keeping track of unit cell coordinates). 
The use of periodic boundary conditions in this manner allows us to use realistic sizes of sphere packings, and still being able to track walkers for long times and average dynamics over local disorder realization in an efficient way. Still, the size of the unit cell is made large enough to render effects of periodicity negligible (sizes of utilized sphere packings are carefully stated). We base our investigations on equilibrium starting conditions, i.e. allowing walkers to start randomly over the whole unit cell (including the interior of the non-scattering regions). For step statistics, only full steps between scattering events are considered. To be on the optical time scale, walkers are set to travel at the speed of $v=200~\upmu$m/ps  (corresponding to a refractive index of 1.5).

% A MONODISPERSIVE SPHERE PACKING
\subsection{A monodisperse sphere packing}\label{subsec:Monodispersive}

\begin{figure}[ht]
  \includegraphics[]{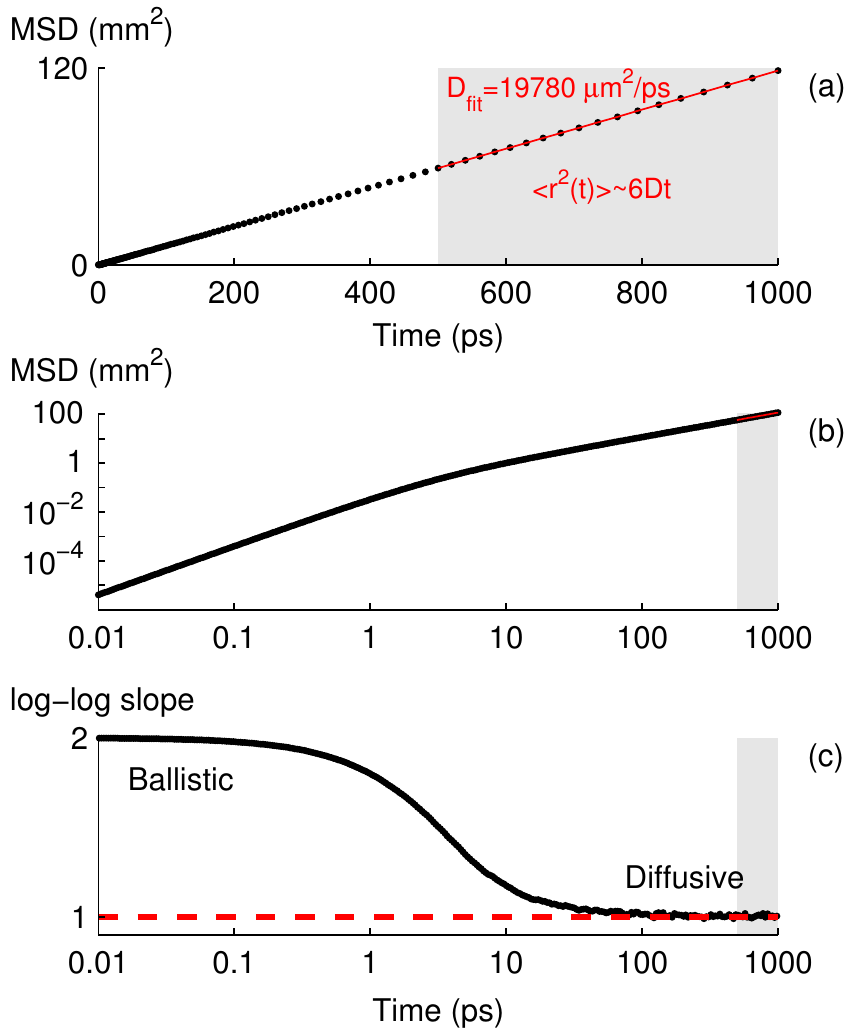}\\
  \caption{Dynamics for a holey system based on a monodispersive sphere packing ($r=50~\upmu$m, $\phi=0.3$, $\ell_s=200~\upmu$m and $v=200~\upmu$m/ps). The diffusion constant extracted from analysis of the MSD evolution is in excellent agreement with the diffusion constant predicted by our theory (Eq. \ref{eq:D_holey}). For this system theory predicts a diffusion constant of $19798~\upmu$m$^2/$ps, and linear fit over the temporal range indicated in gray in part (a) gave $19780\pm 34~\upmu$m$^2/$ps (mean $\pm$ standard error of the mean, as obtained from five simulation repetitions with $10^5$ random walkers each). The middle graph (b) shows a log-log plot of (a). By showing the slope of the log-log plot, the bottom graph (c) reveals the transition from ballistic to diffusive dynamics and verifies that the temporal range used in our analysis falls within the asymptotic diffusive limit.}\label{FIG4_Monodisp_MSD}
\end{figure}

The first case study is a holey system based on a monodispersive sphere packing. The non-scattering part of the material is constituted by randomly placed spheres of radius $50~\upmu$m, the sphere filling fraction being $\phi=0.3.$ Fig. \ref{FIG4_Monodisp_MSD} exemplifies the mean square displacement obtained
by averaging the dynamics of $5\times10^5$ random walkers launched randomly in a cubic system with a side of about 12 mm (1025017 randomly placed spheres). The average spacing between spheres is, in this case, $155.6~\upmu$m (as given by Eq. \ref{eq:Mean_Dss_2}). As discussed in the figure caption, the outcome is in excellent agreement with our theory.

\begin{figure}[h]
  \includegraphics[]{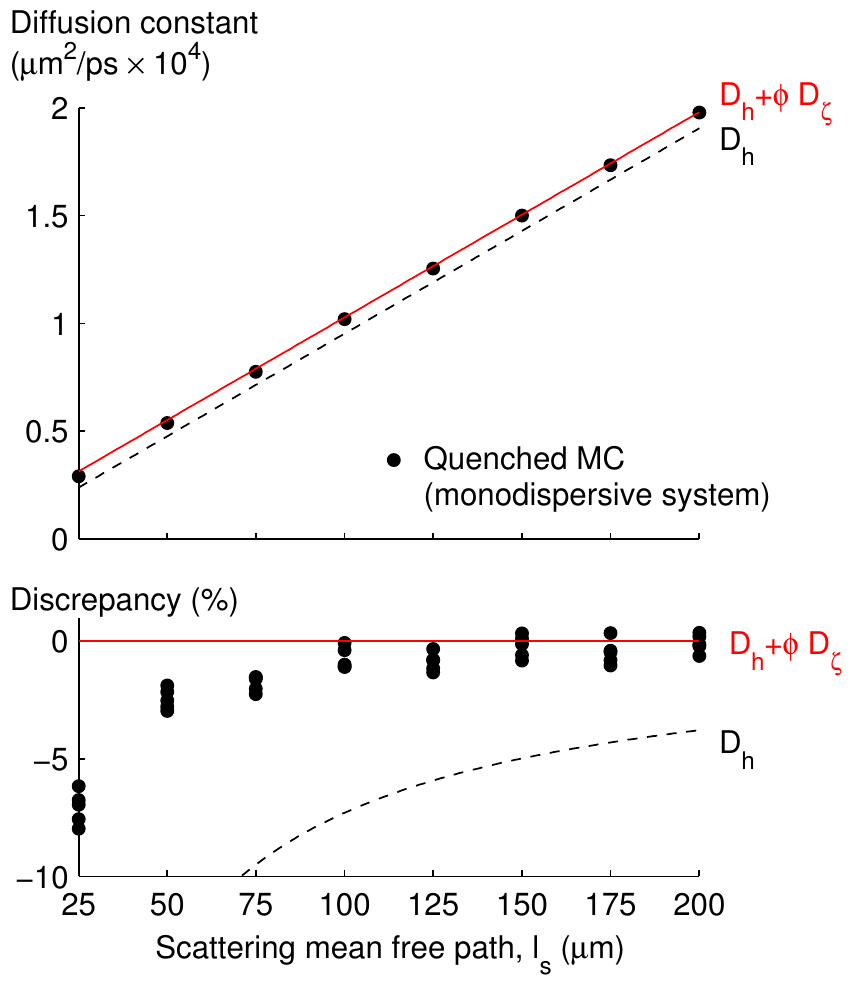}\\
  \caption{The analytical theory for the diffusion constant (red solid line, Eq. \ref{eq:D_holey}) is in excellent agreement with quenched Monte Carlo simulation of transport in a monodispersive sphere packing with $r=50~\upmu$m and $\phi=0.3$ (solid dots, average of 5 repetitions of $10^5$ walkers per level of $\ell_s$). This graph shows how $D$ obtained from simulations exceed the homogenized diffusion constant $D_h$ (dashed line) exactly by the amount $\phi D_\zeta$, largely independent of $\ell_s$. As scattering becomes stronger and stronger (reduction in $\ell_s$), anti-correlations between steps should reduce $D$, and deviations from our theory is expected. The bottom graph clarifies that, in this particular case, this effect becomes important only for $\ell_s$ smaller than approximately $r/2=25~\upmu$m (here, all five repetitions are shown). Note that the mean spacing between spheres is $E[\Dss]\approx 156~\upmu$m.}\label{FIG5_Monodisp_D_vs_ls}
\end{figure}

The applicability of our theory is further elaborated in Fig. \ref{FIG5_Monodisp_D_vs_ls}, in which we study how the diffusion constant depend on the inter-sphere scattering mean free path $\ell_s$. 
Interestingly, the example shows that even for cases where the average spacing between spheres, $E[\Dss]$, is several times larger than $\ell_s$, step correlations are negligible and our analytical theory is applicable. Of course, when $\ell_s$ becomes significantly smaller than the void size, the increase in return probabilities should affect transport (steps become anti-correlated). This in good agreement with the result that step correlations in this monodispersive system is important only for $\ell_s<r/2=25~\upmu$m. The strong quenching regime requires separate treatment and is beyond the scope of this article.

% A FRACTAL SPHERE PACKING
\subsection{A fractal sphere packing}

Let us now turn to the more general case of a polydispersive sphere packing, i.e. a situation where the non-scattering regions vary in size. We will study the limiting case of a system with fractal heterogeneity (over two orders of magnitude) and a high sphere filling fraction. More specifically, our sphere packing is based on $M=18$ sphere categories where the radii $r_i$ are sampled exponentially sampled from $r_\textrm{min}=2.5~\upmu$m up to $r_\textrm{max}=200~\upmu$m, i.e.
\begin{align}\label{eq:r_i_selection}
    r_i = r_\textrm{min} \times \exp\Bigg( \log\bigg(\frac{r_\textrm{max}}{r_\textrm{min}}\bigg) \times \frac{i-1}{M-1}  \Bigg)
\end{align}
The number density of the spheres is set proportional to $1/r_i^{3}$ (i.e.,  the different sphere categories all occupy the same volume fraction), and the filling fraction of the system as a whole is $\phi=0.7005$ (9507676 spheres in a cube with a side of about 2 mm). 

\begin{figure}[h]
  \includegraphics[]{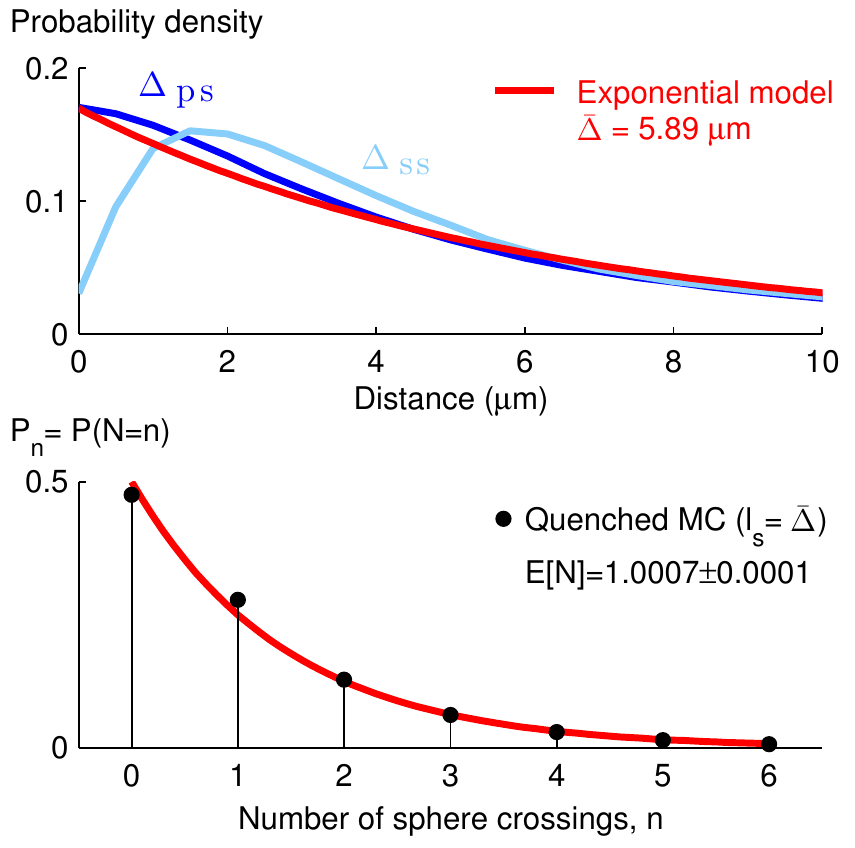}\\
  \caption{The theoretical predictions of  the exponential spacing model is in good agreement with both quenched simulations and statistical analysis of a strongly polydispersive sphere packing. The success is here illustrated for the case of a fractal sphere packing with $M=18$ radius categories ranging from $r_\textrm{min}=2.5~\upmu$m to $r_\textrm{max}=200~\upmu$m (9507676 spheres in a cube with a side of about 2 mm, $\phi=0.7005$). As shown in the top graph, the analytically predicted mean spacing of $\bar{\Delta}=5.89~\upmu$m is in good agreement with spacing distributions obtained via statistical analysis of the sphere packing ($E[\Dps] \approx E[\Dss] \approx 6~\upmu$m). The bottom graph highlights that the crossing probabilities $P_n$ predicted by the exponential spacing model (Eq. \ref{eq:P_n}) agree almost perfectly with simulations running at $\ell_s=\bar{\Delta}$. In particular, setting $\ell_s=\bar{\Delta}$ (as done in this simulation) clearly renders $E[N]=1$. The mean and mean squared steps are also in good agreement with theory. Theory predicts $E[S]=\ell_h=19.669~\upmu$m and $E[S^2]=1776.7~\upmu$m$^2$, and simulations gave $19.679\pm0.004~\upmu$m and $1692\pm1~\upmu$m$^2$, respectively (mean $\pm$ standard error). That the observed mean squared step is a few percent lower than the theoretical prediction is consistent with the fact that the exponential model slightly overestimates the number of higher order multi-crossings ($n \geq 3$). This, in turn, can be understood by looking in the upper graph and noting that the sphere-sphere spacing distance often will be underestimated by the exponential spacing model.}\label{FIG6_Polydisp_ExpSpacing}
\end{figure}

Fig. \ref{FIG6_Polydisp_ExpSpacing} highlights the success of the exponential spacing model also for such a complex heterogeneous system. In particular, this means that we have good quantitative knowledge on the step length distribution (including the sphere crossing statistics). 

\begin{figure}[ht]
  \includegraphics[]{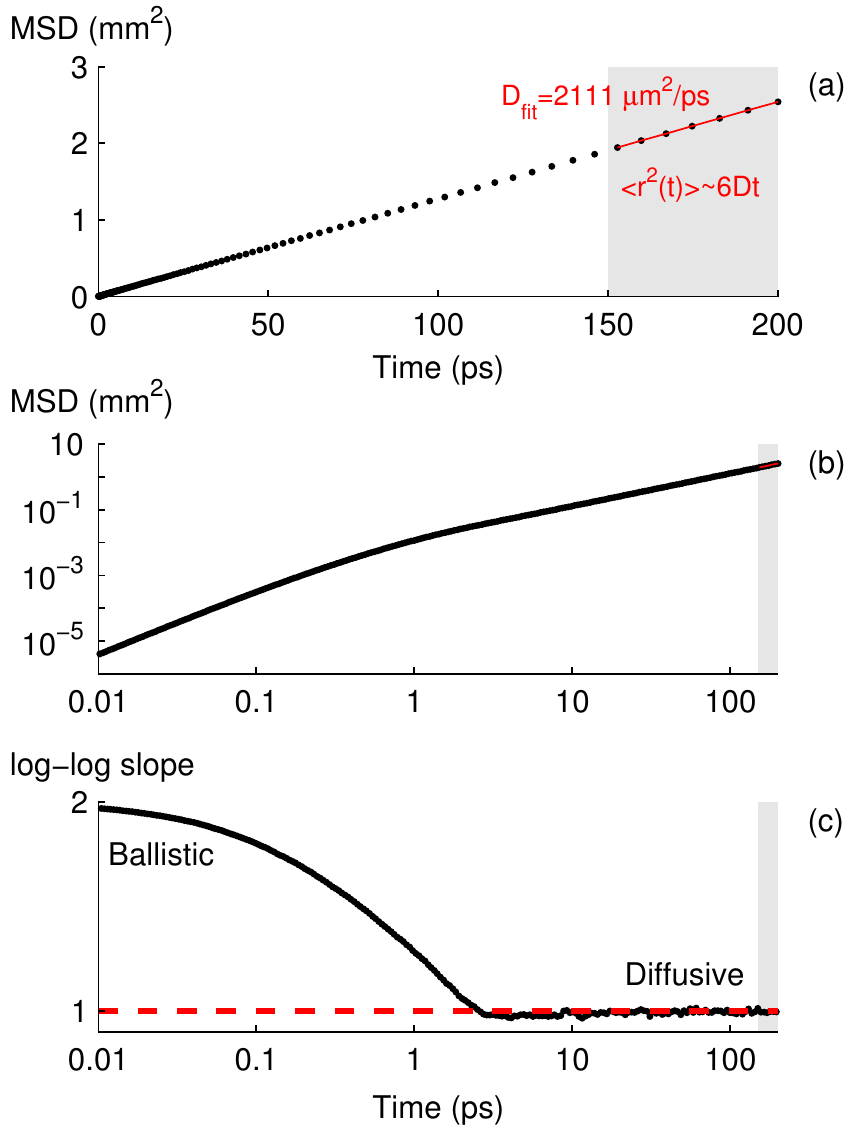}\\
  \caption{Transport dynamics for a holey system based on a fractal sphere packing (sphere radii ranging from $2.5~\upmu$m to $200~\upmu$m, each category occupying about 4\% of the volume, the overall sphere filling fraction being $\phi=0.7005$). Here, the walker crosses on average one sphere in-between scattering events ($\ell_s=\bar{\Delta}=5.89~\upmu$m). The exact $E[S^2]$ as obtained from simulations would, if steps were independent, result in a diffusion constant of $2866~\upmu$m$^2/$ps (step data presented in Fig. \ref{FIG6_Polydisp_ExpSpacing}, but note that exponential spacing model in this particular case overestimates the $E[S^2]/E[S]$ ratio by about 5\%). In contrast, a linear fit over the temporal range indicated in gray in part (a) gives $D=2111\pm 7~\upmu$m$^2/$ps (mean $\pm$ standard error of the mean, as obtained from five simulation repetitions with $10^5$ random walkers each). That the observed $D$ is about 25\% lower is clear evidence that step correlations play an important role. }\label{FIG7_Polydisp_MSD}
\end{figure}

\begin{figure}[ht]
  \includegraphics[]{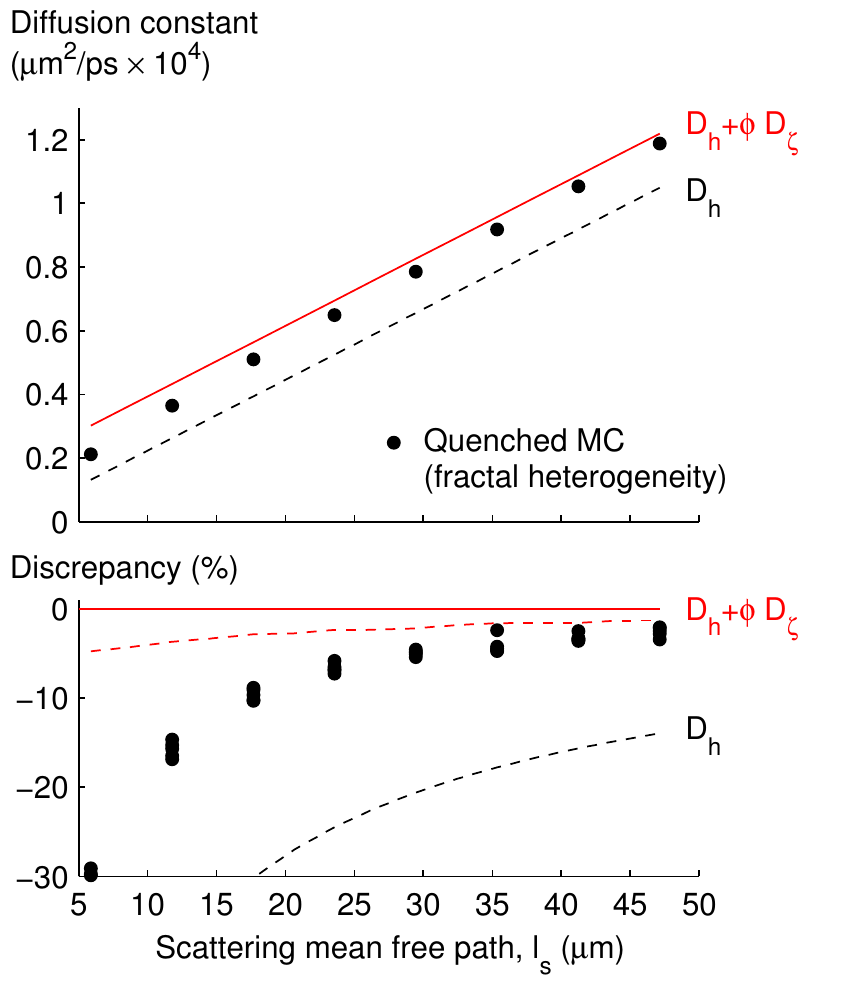}\\
  \caption{Also for strongly polydispersive systems, our probabilitisic theory is successful. The role of step correlations is, however, a complicated matter. In our particular fractal system, where $400~\upmu$m diameter spheres are the largest non-scattering regions, the role of correlation becomes small when $\ell_s$ is a few times larger than the mean void spacing $\bar{\Delta}$. At $\ell_s=4\bar{\Delta}\approx 24~\upmu$m, the asymptotic $D$ obtained from MSD evolution is then only 4\% lower than what the step statistics from quenched simulations tells us. The mean step $E[S]=\ell_h$ is then about $80~\upmu$m, a length which apparently is large enough to render step anti-correlation related to crossing of large spheres relatively small. Here, it should be noted that the exponential spacing model overestimates $E[S^2]$ by a few percent (cf. discussion in Fig. \ref{FIG6_Polydisp_ExpSpacing}). The exact diffusion constant we would see if steps were independent can be estimated from the steps statistics of simulations (via Eq. \ref{eq:Diffusion_constant}), and is in this graph indicated by the dashed red line. Although the errors caused by our model ($D_h+\phi D_\zeta$) is only a few percent, it is still important to realize that step correlation are negligible when the actual diffusion constants (solid dots) meet the dashed line.}\label{FIG8_Polydisp_D_vs_ls}
\end{figure}

Using $\ell_s=\bar{\Delta}$ as a first test case, Fig. \ref{FIG7_Polydisp_MSD} shows the onset of diffusion and the asymptotic diffusion constant. Interestingly, the asymptotic diffusion constant is in this case about 25\% lower than what we would have if steps were independent. With decreasing scattering strength, the importance of step correlations steadily decrease, and the diffusion constant becomes in good agreement with our analytical expression $D=D_h+\phi D_\zeta$ . This behavior is illustrated in Fig. \ref{FIG8_Polydisp_D_vs_ls}, which shows simulations for $\ell_s$ being 1 to 8 times $\bar{\Delta}$. In this context, we want to emphasize that step correlations \textit{per se} do not affect the applicability of the exponential spacing model and the accuracy of the related $E[S^2]$ estimation (Eq. \ref{eq:SSq_summary}). Instead, they only affect the validity of the assumption of independent steps that leads to $D=D_h+\phi D_\zeta$ (Eq. \ref{eq:D_holey}).

Before ending this section, we would like to make a brief comparison to the monodispersive system studied in the previous section. There, we found that step anti-correlations started to play a role only when the scattering mean free path was smaller than the size of the involved spheres. For our polydispersive system, the situation is more complicated. The average spacing between spheres are only a few $\upmu$m, and the sphere sizes range from 5 to 400 $~\upmu$m. One can expect that large spheres will give rise to strong step anti-correlations, but since the transport between these larger spheres is a complicated function also of the smaller spheres in-between, it is difficult to know for which $\ell_s$ this effect becomes important. As shown in Fig. \ref{FIG8_Polydisp_D_vs_ls}, step correlation is of little important once $\ell_s$ is larger than, say, $4\bar{\Delta}\approx 25~\upmu$m. As mentioned also in the figure caption, this scattering strength gives a mean step $E[S]=\ell_h$ of about $80~\upmu$m. That correlations are weak can then be (partly) understood by realizing that most spheres are significantly smaller than this mean step length (the mean chord of the system is, in fact, not longer than $E[\zeta]=13.8~\upmu$m).

\clearpage

% FINITE SIZE FRACTAL SPHERE PACKINGS
\subsection{Finite size fractal sphere packings: L\'evy Glass}

So far, we have investigated unbounded holey systems. In practice, one often deals with finite size system where transport can be even more complicated to understand. For homogenous media it is, for example, well known that the transport in slabs deviates from diffusion when sample thickness $L$ is less than about 10 times the transport mean free path \cite{Yoo1990_PRL,Kop1997_PRL,Elaloufi2004_JOSAA}. For heterogeneous systems, we anticipate that  additional complications arise when the sizes of non-scattering regions are on the order of the sample size. This section therefore aims at discussing the relation between transport in unbounded media, as treated in previous sections, and transport in finite size media. We will show that the theory of diffusion constants is not directly applicable to bounded systems when the size of heterogeneities is on the order of system dimensions.

This issue of bounded heterogeneous systems is particularly relevant to the ongoing discussion on anomalous transport in L\'evy Glass \cite{Barthelemy2008_Nature,Burioni2010b_PRE,Barthelemy2010_PRE,Bertolotti2010_AdvFunctMater,Buonsante2011_PRE,Burresi2012_PRL,Groth2012_PRE}. L\'evy Glasses are turbid materials with strong (fractal) spatial heterogeneity in the density of scatterers. The heterogeneity is engineered and controlled by  the embedding non-scattering regions that follow a power law size distribution into a turbid medium. More specifically, the non-scattering regions are constituted by glass spheres of sizes that are (ideally) exponentially sampled from some smallest radius $r_\textrm{\scriptsize min}$ to an upper radius $r_\textrm{\scriptsize max}$. Assuming one sphere crossing per step and a negligible contribution from the inter-sphere media, setting the number density of spheres $n\sim r^{-(\alpha+2)}$ should produce a step length distribution that fall off as $\ell^{-(\alpha+1)}$ \cite{Bertolotti2010_AdvFunctMater}. On scales on the order of the maximal sphere size, as in slabs of thickness $L=2r_\textrm{\scriptsize max}$, the transport is then believed to be close to that of a L\'evy walk. Although the probabilistic theory presented here represents a significant contribution to the topic of L\'evy Glass design, this falls outside the main focus of this article. A view on this specific but important matter is, nonetheless, given in Appendix \ref{app:LevyGlass}. Here, we will instead stick to the more general question of the relation between transport in unbounded and bounded media. 

The system studied in the previous section is, in fact, an unbounded L\'evy Glass with $\alpha=1$. Its composition closely resembles the L\'evy Glass systems that have been studied experimentally \cite{Barthelemy2008_Nature,Burresi2012_PRL} and simulation-wise \cite{Groth2012_PRE}. Our study has shown that step correlations play an important role for transport in the unbounded systems, and that the resulting diffusion constant for the system was about $2111~\upmu$m$^2$/ps. Looking back at Fig. \ref{FIG7_Polydisp_MSD} and the evolution of the mean square displacement (MSD), it appears that the onset of diffusion occurs surprisingly early. The MSD starts to grow approximately linear with time already after 2 ps. At this time, the MSD is about 0.027 mm$^2$, meaning that the average walker displacement is (roughly) on the order of 150 $\upmu$m. From this, it may be tempting to conclude that transport is diffusive after some 150$~\upmu$m, and that the transmission through a L\'evy Glass of thickness $L=2r_\textrm{max}=400~\upmu$m indeed will be diffusive. In fact, Groth et al. \cite{Groth2012_PRE} have claimed that transport through L\'evy Glass follows regular diffusion. To check whether the above reasoning is adequate or not, and to investigate the claim of Groth et al., we have performed simulations of the transport through a bounded version of this sphere packing. 

\begin{figure}[h]
  \includegraphics[]{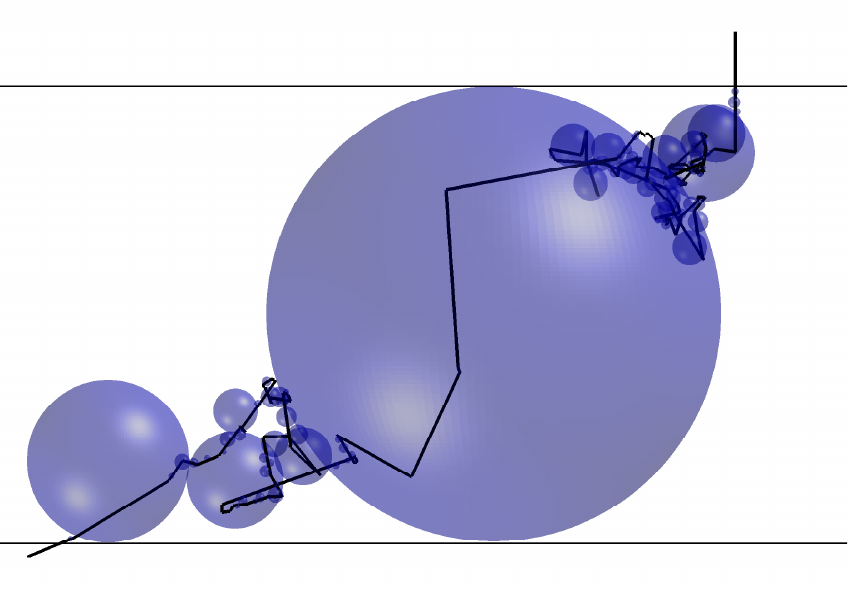}\\
  \caption{2D projection of a simulated trajectory through a 3D L\'evy Glass slab. In this case, the random walker crossed one of the largest spheres. To obtained average transmission properties, walkers are injected randomly on top of the slab (the slab consists of 9507676 spheres, the thickness being $L=2r_\textrm{max}+\epsilon=402~\upmu$m, $\phi=0.7005$). Analysis of the sphere packing reveals that the exponential spacing model applies equally well as in the unbounded case, and that $\ell_s=\bar{\Delta}=5.89~\upmu$m indeed renders $E[N]$ very close to one.}\label{fig:FIG9_LevyGlass_RW}
\end{figure}

\begin{figure}[h]
  \includegraphics[]{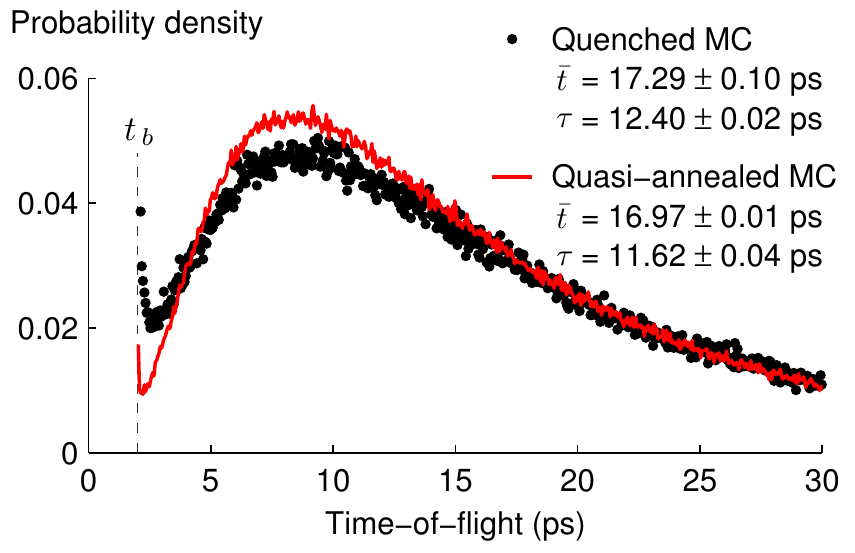}\\
  \caption{Transmission dynamics for a L\'evy Glass, i.e. a bounded fractal sphere packing (slab thickness $L=402~\upmu$m, maximal sphere diameter $400~\upmu$m). Important characteristics such as mean transmission time $\bar{t}$ and decay constant $\tau$ show that the dynamics is not consistent with diffusion (see main text). This conclusion is also supported by the significant amount of quasi-ballistic transmission (truly ballistic component at $t_b=L/v$ is not part of the shown distribution). The dynamics is instead well reproduced by a quasi-annealed simulation based on the exponential spacing model. This mutual agreement indicates that step correlations play a less important role here than in the unbounded case. Statistics are based on 3 and 10 repetitions of $10^6$ walkers for the quenched and quasi-annealed simulations, respectively (mean times and decay constant are stated in the graph as the mean $\pm$ standard error). The transmission were in both cases about 12\%, while the the ballistic fraction of the transmission was 0.6\% and 0.15\% (higher in the quenched case).}\label{fig:FIG10_LevyGlassDynamics}
\end{figure}

Fig. \ref{fig:FIG9_LevyGlass_RW} illustrates our simulations of transport through bounded systems. Our conclusion from these simulations is that transport through L\'evy Glass cannot be described by diffusion. Along with a significant amount of ballistic and quasi-ballistic transmission, we observe major deviations from diffusion theory also in mean transmission time and long-time decay constant (lifetime). Diffusion theory of light transport is well established, and in the diffusive regime it is, for example, well known \cite{Vellekoop2005_PRE} that the mean transmission time $\bar{t}_D$ follows
\begin{align}\label{eq:mean_TOF_D}
  \bar{t}_D = \frac{(L+2z_e)^2}{2D}
\end{align}
and that the decay time constant $\tau_D$  is given by
\begin{align}\label{eq:tau_D}
  \tau_D = \frac{(L+2z_e)^2}{\pi^2D} 
\end{align}
In the above formulas, $L$ denotes slab thickness, and $z_e=\frac{2}{3}\ell_t$ the so called extrapolation length. If transport through the slab is diffusive with $D=2111~\upmu$m$^2$/ps, macroscopic transport should be identical to a random walk of independent and exponential distributed steps with average length $\ell_t=3D/v\approx 33~\upmu$m. We would thus expected a mean transmission time of about $\bar{t}_D=15$ ps and a decay constant of about $\tau_D=9$ ps. In contrast, as shown in Fig. \ref{fig:FIG10_LevyGlassDynamics}, quenched simulations result in a mean transmission time of 17.4 ps and decay constant of 12.5 ps. In fact, the observed combination of $\bar{t}$ and $\tau$ is in fact not consistent with any diffusion constant. Furthermore, the significant amount of ballistic and quasi-ballistic light alone indicates that diffusion cannot well capture the transport. We therefore conclude that transport in L\'evy Glass is not governed by diffusion. Instead, we found that the dynamics is well reproduced by, what we call, a quasi-annealed simulation in which we mimic the step distribution but effectively remove step anti-correlations. In this straight-forward simulation, being a numerical correspondent to our probabilistic theory, steps are generated on the basis of the exponential spacing model under the requirement that when the walker is about to a make a chord, this chord cannot be longer than the distance to the slab boundary (see  Appendix \ref{app:QuasiAnnealed} for a more detailed description). The step length is thus position and direction dependent, but step correlations are heavily suppressed. Although it is out of scope of this article to investigate this interesting result in detail, it appears as if the presence of boundaries in the quenched system strongly alter the role of step correlations.

\section{Conclusions}

We have presented a probabilistic theory for transport in systems with quenched heterogeneous distribution of scatterers. The focus lied on random systems, and via a simple model of the quenched disorder we have been able to derive analytical expressions for, notably, the mean step, the mean squared step, the diffusion constant and void crossing probabilities. Here we repeat two key results. First, we have shown that the mean step is given by (Eq. \ref{eq:E_S})
\begin{align}
   E[S] = \frac{\ell_s}{1-\phi} = \ell_h \nonumber
\end{align}
and thus equals the homogenized (exponential) scattering mean free path $\ell_h$, being dependent only of void filling fraction $\phi$, not their size and shape. Second, for holey systems in the form of random sphere packings we have also shown (in the limit of weak step correlations) that the asymptotic diffusion constant is (Eq. \ref{eq:D_holey})
\begin{align}
   D = \frac{1}{3} v \ell_h + \phi \times \frac{1}{6}v \times \frac{E[\zeta^2]}{E[\zeta]} \nonumber
\end{align}
In cases where step correlations are non-negligible, under the reasonable assumption that steps are not positively correlated, this expression can function as an upper bound on the diffusion constant. All our theoretical results are well supported by Monte Carlo simulations of random walks in holey systems based on random sphere packings. We have also shown that the relation between transport in unbounded and bounded system are particularly complex when heterogeneities are on the order the system size (especially regarding the onset of regular diffusion and the role of step correlations).

Given the abundance of turbid materials where scatterers are not homogeneously distributed, we believe that both our viewpoint and our results can be of value for a wide range of topics \--- from fundamental work on light transport to applied spectroscopy of heterogeneous media such as biological tissues, food products, and powder compacts. In particular, viewing transport as a quenched holey random walk may turn out to be an important approach for work directed towards the understanding of the optics and spectroscopy of complex porous media with broad and/or anisotropic pore size distribution, such as pharmaceutical tablets \cite{Svensson2010_OptLett,Alerstam2012_PRE}. In this context, it should be noted that generalization of our work to systems with impedance mismatch and anisotropic heterogeneity should be rather straight forward. Account for refractive index mismatch is, for example, standard in the study of light transport  \cite{Wang1995_CMPB,Sadjadi2011_PRE}. Finally, as discussed earlier, a theoretical account of quenched systems with strong step correlations is more challenging, but equally important.

\begin{acknowledgments}
This work benefitted from the European Network of Excellence for Nanophotonics for Energy Efficiency and the ERC grant Photbots. T.S. gratefully acknowledge funding from the Swedish Research Council. Erik Alerstam, Raffaella Burioni and Alessandro Vezzani are thanked for useful discussions.
\end{acknowledgments}

% APPENDICES
\appendix

% E[S^2] DERIVATIONS
\section{Derivations for $E[S^2]$}\label{app:E_SSq}

In this appendix, we provide a derivation of the expression for the mean squared step  given in the main text. We start by expanding the mean squared step into three components:
\begin{align}\label{eq:S_sq_expansion}
   E[S^2] =&\quad E[S^2_\textrm{turbid}] + E[S^2_\textrm{void}] + 2E[S_\textrm{turbid}S_\textrm{void}]
\end{align}
Since $E[S^2_\textrm{turbid}]=2\ell_s^2$ per definition, the challenge is the last two terms. As mentioned in the main text, we will make the important assumption that the length of the different chords involved in a step is independent on the number of spheres crossed (this means, e.g., that we assume that the spacing between two spheres are independent of the sizes of these spheres). By splitting the outcome of $S_\textrm{void}$ into cases of different number of sphere crossings, we can calculate the two terms from the rules of total expectation values. Starting with $E[S^2_\textrm{void}] $, we have
\begin{align}\label{eq:S_void_Sq}
   E[S^2_\textrm{void}] &= \sum\limits_{n=0}^{\infty} P_n \times E[S^2_\textrm{void} | N=n] \nonumber \\
           &= \sum\limits_{n=0}^{\infty} P_n \times E [ (\sum\limits_{i=0}^{n} \zeta_i)^2 ] = \nonumber \\
           &= \sum\limits_{n=0}^{\infty} P_n \times (nE[\zeta^2] + n(n-1)E[\zeta]^2) \nonumber \\
           &= E[N]E[\zeta^2] + \bigg(E[N^2]-E[N]\bigg)E[\zeta]^2 \nonumber \\
           &= \phi\ell_h\frac{E[\zeta^2]}{E[\zeta]} + 2\frac{\ell_s^2}{\bar{\Delta}^2}E[\zeta]^2
\end{align}

When calculating $E[S_\textrm{turbid}S_\textrm{void}]$, we must remember that they are dependent variables. We can break this dependence by, again, splitting into cases of different number of sphere crossings
\begin{align}\label{eq:Sturbid_x_Ssphere}
   E[S_\textrm{turbid}S_\textrm{void}] &= \sum\limits_{i=0}^\infty P_n\times E[S_\textrm{turbid}S_\textrm{void} | N_\textrm{cross}=n] \nonumber \\
   &=  \sum\limits_{i=0}^\infty P_n\times nE[\zeta] \times E[S_\textrm{turbid} | N=n] \nonumber\\
   &= \sum\limits_{i=0}^\infty P_n\times nE[\zeta] \times(n+1) \frac{\bar{\Delta}\ell_s}{\bar{\Delta}+\ell_s} \nonumber \\
   &= E[\zeta] \frac{\bar{\Delta}\ell_s}{\bar{\Delta}+\ell_s}  \times \bigg(E[N^2]+E[N]\bigg) \nonumber \\
   &= 2E[\zeta] \frac{\bar{\Delta}\ell_s}{\bar{\Delta}+\ell_s}  \times \bigg(E[N]^2+E[N]\bigg) \nonumber \\
   &= 2E[\zeta] \frac{\bar{\Delta}\ell_s}{\bar{\Delta}+\ell_s}  \times \bigg(\frac{\ell_s^2}{\bar{\Delta}^2}+\frac{\ell_s}{\bar{\Delta}}\bigg) \nonumber \\
   &= 2E[\zeta] \frac{\bar{\Delta}\ell_s}{\bar{\Delta}+\ell_s}  \times \frac{\ell_s(\ell_s+\bar{\Delta})}{\bar{\Delta}^2} \nonumber \\
   &= 2E[\zeta] \times \frac{\ell_s^2}{\bar{\Delta}} \nonumber \\ 
\end{align}
The step to reach the third line, involving a calculation of $E[S_\textrm{turbid} | N=n]$, deserves a comment. This term tells us the average  inter-void path made given a certain number of crossings. Having $n$ crossings is the same as knowing that the step was longer than a sum of $n$ spacings, but that the step in the could not cross the $(n+1)$:th spacing. This event can be written as
\begin{align}\label{eq:Event_N_crossings}
    \sum\limits_{i=1}^{n}\Delta_i < S_\textrm{turbid} < (\sum\limits_{i=1}^{n}\Delta_i) + \Delta_{i+1}.
\end{align}
Noting that the sum of the $n$ spacing is an Erlang distribution (sum on $n$ independent exponential distributions), we can calculate $E[S_\textrm{turbid} | N=n]$ by doing three-dimensional integration over the joint distribution of three random variables (the Erlang distribution, $\Delta_{n+1}\in\textrm{Exp}(\bar{\Delta})$ and $S_\textrm{turbid}\in\textrm{Exp}(\ell_s)$). The conditional expectation value is reached by integrating over the space corresponding to the event $N=n$, as expressed in Eq. \ref{eq:Event_N_crossings} (note that the joint density function must be renormalized by the probability of the event). The result is, in essence, the Gamma function integral. A more elegant way to solve the problem is, however, to see the problem as an encounter of spacings $\Delta_i$, where the memoryless nature of the random process makes it possible to treat each and every crossing independently. If a walker manages to cross a certain spacing, the best estimate of the spacing crossed is $E[\Delta | S_\textrm{turbid}>\Delta]$. On the other hand, when the walker finally does \textit{not} manage to cross a spacing, the step made in this last part (after the last void crossing) is on average $E[S_\textrm{turbid} | \Delta>S_\textrm{turbid}]$. These two terms can be calculated by means of a two-dimensional integral over their joint probability density function (the symmetry of the problem makes one term follow from the other).
\begin{align}\label{eq:E_spacing_crossed}
   E[\Delta | S_\textrm{turbid}>\Delta]  &= \frac{\iiint\limits_{\ell>\delta} \delta \times p(\ell,\delta)d\ell d\delta}{\iiint\limits_{\ell>\delta} p(\ell,\delta) \ell d\delta} \nonumber \\
       &= \frac{\bar{\Delta}\ell_s}{\bar{\Delta}+\ell_s} \nonumber \\
       &= E[S_\textrm{turbid} | \Delta>S_\textrm{turbid}]
\end{align}

So, when a walker manages to cross a inter-sphere spacing, that spacing is, on average, $E[\Delta | S_\textrm{turbid}>\Delta]$. As emphasized above, the memoryless nature of exponential distribution allows us to forget the total inter-sphere step already made, and make the same conclusion at after each crossing. When the walker finally fails to cross a spacing, the average step taken after the last void crossing was $E[S_\textrm{turbid} | \Delta>S_\textrm{turbid}]$. Given that there was $n$ crossings in total, the expected path is therefore
\begin{align}\label{eq:E_Turbid_given_N}
    E[S_\textrm{turbid} | N=n] &= n\times E[\Delta | S_\textrm{turbid}>\Delta] \nonumber \\ 
    & \quad + E[S | \Delta>S_\textrm{turbid}]  \nonumber \\
    &=  (n+1) \frac{\bar{\Delta}\ell_s}{\bar{\Delta}+\ell_s}
\end{align}

We reach the expression Eq. \ref{eq:SSq_summary} by summing the three contributions and reducing an appearing quadratic polynomial
\begin{align}\label{eq:S_sq_expansion_2}
   E[S^2] &= E[S^2_\textrm{turbid}] + E[S^2_\textrm{void}] + 2E[S_\textrm{turbid}S_\textrm{void}] \nonumber \\
           &= 2\ell_s^2 + \phi\ell_h\frac{E[\zeta^2]}{E[\zeta]} + 2\frac{\ell_s^2}{\bar{\Delta}^2}E[\zeta]^2 + 4E[\zeta] \times \frac{\ell_s^2}{\bar{\Delta}} \nonumber \\
           &= 2\ell_s^2 \bigg(1+\frac{E[\zeta]}{\bar{\Delta}} \bigg)^2 + \phi\ell_h\frac{E[\zeta^2]}{E[\zeta]} \nonumber \\
           &= 2\ell_s^2 \bigg(1+\frac{\phi}{1-\phi} \bigg)^2 + \phi\ell_h\frac{E[\zeta^2]}{E[\zeta]} \nonumber \\
           &= 2\ell_h^2 + \phi\ell_h\frac{E[\zeta^2]}{E[\zeta]}
\end{align}

% LEVY GLASS DESIGN
\section{L\'evy Glass design}\label{app:LevyGlass}

When  L\'evy Glass were introduced by Barthelemy et al. \cite{Barthelemy2008_Nature}, the scatterer concentration was selected aiming at achieving, on average, one scattering event between sphere crossings (i.e. $E[N]=1$, where $N$ states the number of sphere crossings in a step). The design principle was not elaborated in more detail, and exactly how filling fraction and scattering strength affect crossings statistics was not known. The inter-sphere transport mean free path was in fact chosen to match the smallest sphere diameter, i.e. $\ell_t=2r_\textrm{min}=5~\upmu$m. The probabilistic theory presented in this work allows a selection of the scattering strength that more strictly fulfills the ideal L\'evy Glass design criteria. From Eq. \ref{eq:N_cross_mean_exp_model}, we see that $E[N]=1$ is reached when $\ell_s=\bar{\Delta}$. As can be seen in Fig. \ref{FIG6_Polydisp_ExpSpacing}, this result is confirmed by our simulations. In this respect (using Eq. \ref{eq:MeanSpacing} to express $\bar{\Delta}$), the ideal scattering strength is therefore
\begin{align}\label{eq:Ideal_l_s}
   \ell_s^\textrm{ideal} = \bar{\Delta} = \frac{1-\phi}{\phi}\times E[\zeta] 
\end{align}
Returning to the experimental work in Ref. \cite{Barthelemy2008_Nature}, the filling fraction, as can be calculated from recipe given in the supplemental information, was about 67\%. Via calculation of the mean chord, we find that the ideal $\ell_s$ for the samples used to investigate transmission scaling ranges from $9.5~\upmu$m ($L=550~\upmu$m, $L=2r_\textrm{\scriptsize max}$ being the L\'evy Glass slab thickness) to $4.2~\upmu$m ($L=50~\upmu$m). Although the utilized scattering strength clearly is on the right order of magnitude, setting $\ell_s=2r_\textrm{min}$ is not universally ideal. That the ideal $\ell_s$ varies with L\'evy Glass thickness is related to the fact that thicker samples include the use of larger spheres that fill space more efficiently than smaller ones. If the filling fraction $\phi$ is kept constant, the average sphere spacing will therefore increase with thickness. It is thus important to realize that studies of a series of L\'evy Glass with varying $L=2r_i$ but with constant $\phi$ and $\ell_s$ comes with an inherent mismatch with the ideal L\'evy Glass design principle. The implication this finding has on the design of scaling experiments remains to be elaborated. For completeness, it should also be noted that scattering is not fully isotropic in the discussed experiments. The titania nanoparticles used gives a anisotropy factor of $g\approx 0.6$, meaning that $\ell_s=(1-g)\ell_t \neq\ell_t$. This is an additional aspect that should be considered in the future. 

In a more recent experimental paper, Burresi et al. \cite{Burresi2012_PRL} reported on weak localization of light (coherent backscattering) in L\'evy glass. There, samples had a filling fraction of about 70\% and were manufactured using spheres of sizes from $r_\textrm{min}=2.5~\upmu$m up to $r_\textrm{max}=115~\upmu$m ($L\approx230~\upmu$m). Our theory shows that the ideal $\ell_s$ is around $6.3~\upmu$m and $3.8~\upmu$m for the studied $\alpha=1$ and $\alpha=1.5$ samples, respectively. In the article, reference experiments where scatterers were dispersed homogeneously rendered $\ell_t=19~\upmu$m, indicating that the inter-sphere scattering strength was $l_t=19/(1-\phi)=5.7~\upmu$m. The experiments done were thus in close agreement with the ideal design principle (again, disregarding that scattering was anisotropic with $g=0.6$).

Finally, a recent paper by Groth, Akhmerov and Beenakker \cite{Groth2012_PRE} investigated transmission scaling in L\'evy Glasses via simulations of random walks in sphere packings. There, simulations are stated to run with $\ell_s=r_\textrm{min}/2$ and it is said that this choice was made to ensure that "there is, on average, one scattering event between leaving and entering a sphere" (i.e., to ensure $E[N=1]$). Given the findings of our present work, we believe that this level of scattering is too strong to render $E[N]=1$ (at least for relevant filling fractions). Above, we reported that experimental L\'evy Glass systems have had a filling fraction of about 70\%, and that the ideal scattering mean free path always is significantly larger than the radius of the smallest sphere. Assuming that Groth et al. are not studying very different systems, we believe that using $\ell_s=r_\textrm{min}/2$ brings them far from the ideal situation where $E[N]=1$. However, since utilized filling fractions were not stated, we cannot be fully quantitative. Nonetheless, it is important to note that while setting $\ell_s$ very small will take us close to the desired $\ell^{-(\alpha+1)}$ power law decay in the step length distribution (multi crossing getting increasing unlikely, i.e., $P(N>1)$ negligible), such a procedure will induce strong step correlations that affect transport.

% THE QUASI-ANNEALED RANDOM WALK
\section{The quasi-annealed random walk}\label{app:QuasiAnnealed}

The proposed quasi-annealed random walk aims at mimicking the step length distribution of the quenched system while removing step correlations related to the quenched disorder. After random walker initiation, and after each scattering event, a random step $S_\textrm{turbid}\in \textrm{Exp}(\ell_s)$ to be travelled through the turbid component is generated. The length of this step is first compared to the distance to the boundary, $\Delta_b$, along the current direction of propagation. If $S_\textrm{turbid}>\Delta_b$, the walker leaves the sample and a transmission or reflection time is registered. If, on the other hand, $S_\textrm{turbid}<\Delta_b$, $S_\textrm{turbid}$ is instead compared to a randomly generated exponential distributed distance to the "closest virtual sphere", $\Delta_s \in \textrm{Exp}(\bar{\Delta})$ (cf. the exponential spacing model described in Sect. \ref{sec:exp_spacing_model}). If $S_\textrm{turbid}<\Delta_s$, scattering will take place, and the procedure will start over when a new direction and a new step $S_\textrm{turbid}$ have been generated. If $S_\textrm{turbid}>\Delta_s$, the walker will cross a non-scattering region, after which the remaining part of the step $S_\textrm{turbid}-\Delta_s$ will be used as the new step forward (to be compared with the updated $\Delta_b$ and a new sphere spacing). The length travelled through the void (a random chord $\zeta$) is generated based on the distribution of chords predicted by equilibrium considerations, i.e. the density function given in Eq. \ref{eq:pdf_chord}. The related cumulative distribution functions (CDF) is
\begin{align}\label{eq:cdf_chord}
    F_\zeta(x) &= P(\zeta\leq x)=1-P(\zeta > x) \\
               &= 1 - \sum\limits_{i:2r_i>x} \;\int\limits_x^{2r_i} p_i\times \frac{x}{2r_i^2}\,dx \\
               &= 1 - \sum\limits_{i:2r_i>x} p_i\times\bigg( 1-\frac{x^2}{4r_i^2} \bigg) \\
               &= 1 - \sum\limits_{i:2r_i>x} p_i + \sum\limits_{i:2r_i>x}p_i\times\frac{x^2}{4r_i^2}
\end{align}
Solving $y=F_\zeta(x)$, we find
\begin{align}\label{eq:inverse_chord_cdf}
    x = F^{-1}_\zeta(y) = \sqrt{\frac{y-1+\sum\limits_{i:F(2r_i)>y} p_i}{\sum\limits_{i:F(2r_i)>y}\frac{p_i}{4r_i^2}}}.
\end{align}
We use this inverse CDF to generate the random chord length $\zeta$, setting
\begin{align}\label{eq:inverse_chord_cdf}
    \zeta = F^{-1}_\zeta(U),
\end{align}
where $U$ is a random variable uniformly distributed in $[0,1]$. Note that a generated chord is only accepted if it is smaller than the current distance to the boundary. If a  generated chord is larger than the distance to the boundary it is discarded, and the generation is remade. This means that long steps to some extent are under-represented compared to the equilibrium consideration that lies behind the chord length distribution. At this stage, it cannot be ruled out that this has some impact on quantities such as mean transit time and decay rate (lifetime). The role of step correlations in transport through quenched disorder certainly deserves further attention in the future.

% BIBLIOGRAPHY


\begin{thebibliography}{10}
\expandafter\ifx\csname url\endcsname\relax
  \def\url#1{\texttt{#1}}\fi
\expandafter\ifx\csname urlprefix\endcsname\relax\def\urlprefix{URL }\fi
\providecommand{\bibinfo}[2]{#2}
\providecommand{\eprint}[2][]{\url{#2}}

\bibitem{Thomas2002_Book}
\bibinfo{author}{Thomas, G.} \& \bibinfo{author}{Stamnes, K.}
\newblock \emph{\bibinfo{title}{Radiative Transfer in the Atmosphere and
  Ocean}} (\bibinfo{publisher}{Cambridge University Press},
  \bibinfo{year}{2002}).

\bibitem{Mishchenko2006_Book}
\bibinfo{author}{Mishchenko, M.}, \bibinfo{author}{Travis, L.} \&
  \bibinfo{author}{Lacis, A.}
\newblock \emph{\bibinfo{title}{{Multiple scattering of light by particles:
  radiative transfer and coherent backscattering}}}
  (\bibinfo{publisher}{Cambridge University Press}, \bibinfo{year}{2006}).

\bibitem{Davis2010_RepProgPhys}
\bibinfo{author}{Davis, A.~B.} \& \bibinfo{author}{Marshak, A.}
\newblock \bibinfo{title}{Solar radiation transport in the cloudy atmosphere: a
  3{D} perspective on observations and climate impacts}.
\newblock \emph{\bibinfo{journal}{Rep. Prog. Phys.}}
  \textbf{\bibinfo{volume}{73}}, \bibinfo{pages}{026801}
  (\bibinfo{year}{2010}).

\bibitem{Tuchin2007_Book}
\bibinfo{author}{Tuchin, V.~V.}
\newblock \emph{\bibinfo{title}{Tissue Optics: Light Scattering Methods and
  Instruments for Medical Diagnosis}} (\bibinfo{publisher}{SPIE/International
  Society for Optical Engineering}, \bibinfo{year}{2007}).

\bibitem{Welch2010_Book}
\bibinfo{author}{Welch, A.} \& \bibinfo{author}{Gemert, M.}
\newblock \emph{\bibinfo{title}{{Optical-Thermal Response of Laser-Irradiated
  Tissue}}} (\bibinfo{publisher}{Springer}, \bibinfo{year}{2010}),
  \bibinfo{edition}{2} edn.

\bibitem{Wang2007_Book}
\bibinfo{author}{Wang, L.} \& \bibinfo{author}{Wu, H.}
\newblock \emph{\bibinfo{title}{Biomedical Optics: Principles and Imaging}}
  (\bibinfo{publisher}{Wiley}, \bibinfo{year}{2007}).

\bibitem{Berne2000_Book}
\bibinfo{author}{Berne, B.} \& \bibinfo{author}{Pecora, R.}
\newblock \emph{\bibinfo{title}{Dynamic Light Scattering: With Applications to
  Chemistry, Biology, and Physics}} (\bibinfo{publisher}{Dover Publications},
  \bibinfo{year}{2000}).

\bibitem{Reich2005_AdvDrugDeliverRev}
\bibinfo{author}{Reich, G.}
\newblock \bibinfo{title}{Near-infrared spectroscopy and imaging: Basic
  principles and pharmaceutical applications}.
\newblock \emph{\bibinfo{journal}{Adv. Drug Deliver. Rev.}}
  \textbf{\bibinfo{volume}{57}}, \bibinfo{pages}{1109--1143}
  (\bibinfo{year}{2005}).

\bibitem{Siesler2008_Book}
\bibinfo{author}{Siesler, H.}, \bibinfo{author}{Ozaki, Y.},
  \bibinfo{author}{Kawata, S.} \& \bibinfo{author}{Heise, H.}
\newblock \emph{\bibinfo{title}{Near-Infrared Spectroscopy: Principles,
  Instruments, Applications}} (\bibinfo{publisher}{John Wiley \& Sons},
  \bibinfo{year}{2008}).

\bibitem{Shi2010a_JPharmSci}
\bibinfo{author}{Shi, Z.~Q.} \& \bibinfo{author}{Anderson, C.~A.}
\newblock \bibinfo{title}{Pharmaceutical applications of separation of
  absorption and scattering in near-infrared spectroscopy ({NIRS})}.
\newblock \emph{\bibinfo{journal}{J. Pharm. Sci.}}
  \textbf{\bibinfo{volume}{99}}, \bibinfo{pages}{4766--4783}
  (\bibinfo{year}{2010}).

\bibitem{Shaw2002_JQSRT}
\bibinfo{author}{Shaw, R.~A.}, \bibinfo{author}{Kostinski, A.~B.} \&
  \bibinfo{author}{Lanterman, D.~D.}
\newblock \bibinfo{title}{Super-exponential extinction of radiation in a
  negatively correlated random medium}.
\newblock \emph{\bibinfo{journal}{J. Quant. Spectrosc. Radiat. Transfer}}
  \textbf{\bibinfo{volume}{75}}, \bibinfo{pages}{13--20}
  (\bibinfo{year}{2002}).

\bibitem{Davis2004_JQSRT}
\bibinfo{author}{Davis, A.~B.} \& \bibinfo{author}{Marshak, A.}
\newblock \bibinfo{title}{Photon propagation in heterogeneous optical media
  with spatial correlations: enhanced mean-free-paths and
  wider-than-exponential free-path distributions}.
\newblock \emph{\bibinfo{journal}{J. Quant. Spectrosc. Radiat. Transfer}}
  \textbf{\bibinfo{volume}{84}}, \bibinfo{pages}{3--34} (\bibinfo{year}{2004}).

\bibitem{Davis2011_JQSRT}
\bibinfo{author}{Davis, A.~B.} \& \bibinfo{author}{Mineev-Weinstein, M.~B.}
\newblock \bibinfo{title}{Radiation propagation in random media: From positive
  to negative correlations in high-frequency fluctuations}.
\newblock \emph{\bibinfo{journal}{J. Quant. Spectrosc. Radiat. Transfer}}
  \textbf{\bibinfo{volume}{112}}, \bibinfo{pages}{632--645}
  (\bibinfo{year}{2011}).

\bibitem{Bal2011_JQSRT}
\bibinfo{author}{Bal, G.} \& \bibinfo{author}{Jing, W.}
\newblock \bibinfo{title}{Fluctuation theory for radiative transfer in random
  media}.
\newblock \emph{\bibinfo{journal}{J. Quant. Spectrosc. Radiat. Transfer}}
  \textbf{\bibinfo{volume}{112}}, \bibinfo{pages}{660--670}
  (\bibinfo{year}{2011}).

\bibitem{Lovejoy1990_JGeophysRes}
\bibinfo{author}{Lovejoy, S.}, \bibinfo{author}{Davis, A.},
  \bibinfo{author}{Gabriel, P.}, \bibinfo{author}{Schertzer, D.} \&
  \bibinfo{author}{Austin, G.~L.}
\newblock \bibinfo{title}{Discrete angle radiative transfer 1. scaling and
  similarity, universality and diffusion}.
\newblock \emph{\bibinfo{journal}{J. Geophys. Res.}}
  \textbf{\bibinfo{volume}{95}}, \bibinfo{pages}{11699--11715}
  (\bibinfo{year}{1990}).

\bibitem{Davis1999_ARM_Proc}
\bibinfo{author}{Davis, A.}, \bibinfo{author}{Marshak, A.} \&
  \bibinfo{author}{Pfeilsticker, K.}
\newblock \bibinfo{title}{Anomalous {L}\'evy photon diffusion theory: toward a
  new parameterization of shortwave transport in cloudy columns}.
\newblock \emph{\bibinfo{journal}{9th ARM Science Team Meeting Proceedings}}
  (\bibinfo{year}{1999}).

\bibitem{Scholl2006_JGeophysRes}
\bibinfo{author}{Scholl, T.}, \bibinfo{author}{Pfeilsticker, K.},
  \bibinfo{author}{Davis, A.~B.}, \bibinfo{author}{Klein~Baltink, H.},
  \bibinfo{author}{Crewell, S.}, \bibinfo{author}{Löhnert, U.},
  \bibinfo{author}{Simmer, C.}, \bibinfo{author}{Meywerk, J.} \&
  \bibinfo{author}{Quante, M.}
\newblock \bibinfo{title}{Path length distributions for solar photons under
  cloudy skies: Comparison of measured first and second moments with
  predictions from classical and anomalous diffusion theories}.
\newblock \emph{\bibinfo{journal}{J. Geophys. Res.}}
  \textbf{\bibinfo{volume}{111}}, \bibinfo{pages}{D12211}
  (\bibinfo{year}{2006}).

\bibitem{Liu1995_MedPhys}
\bibinfo{author}{Liu, H.}, \bibinfo{author}{Chance, B.},
  \bibinfo{author}{Hielscher, A.}, \bibinfo{author}{Jacques, S.} \&
  \bibinfo{author}{Tittel, F.}
\newblock \bibinfo{title}{Influence of blood-vessels on the measurement of
  hemoglobin oxygenation as determined by time-resolved reflectance
  spectroscopy}.
\newblock \emph{\bibinfo{journal}{Med. Phys.}} \textbf{\bibinfo{volume}{22}},
  \bibinfo{pages}{1209--1217} (\bibinfo{year}{1995}).

\bibitem{Firbank1996_PhysMedBiol}
\bibinfo{author}{Firbank, M.}, \bibinfo{author}{Arridge, S.~R.},
  \bibinfo{author}{Schweiger, M.} \& \bibinfo{author}{Delpy, D.~T.}
\newblock \bibinfo{title}{An investigation of light transport through
  scattering bodies with non-scattering regions}.
\newblock \emph{\bibinfo{journal}{Phys. Med. Biol.}}
  \textbf{\bibinfo{volume}{41}}, \bibinfo{pages}{767--783}
  (\bibinfo{year}{1996}).

\bibitem{Hielscher1998_PhysMedBiol}
\bibinfo{author}{Hielscher, A.}, \bibinfo{author}{Alcouffe, R.} \&
  \bibinfo{author}{Barbour, R.}
\newblock \bibinfo{title}{Comparison of finite-difference transport and
  diffusion calculations for photon migration in homogeneous and heterogeneous
  tissues}.
\newblock \emph{\bibinfo{journal}{Phys. Med. Biol.}}
  \textbf{\bibinfo{volume}{43}}, \bibinfo{pages}{1285--1302}
  (\bibinfo{year}{1998}).

\bibitem{Boas2002_OptExpress}
\bibinfo{author}{Boas, D.}, \bibinfo{author}{Culver, J.},
  \bibinfo{author}{Stott, J.} \& \bibinfo{author}{Dunn, A.}
\newblock \bibinfo{title}{Three dimensional {M}onte {C}arlo code for photon
  migration through complex heterogeneous media including the adult human
  head}.
\newblock \emph{\bibinfo{journal}{Opt. Express}} \textbf{\bibinfo{volume}{10}},
  \bibinfo{pages}{159--170} (\bibinfo{year}{2002}).

\bibitem{Bal2002_JCompPhys}
\bibinfo{author}{Bal, G.}
\newblock \bibinfo{title}{Particle transport through scattering regions with
  clear layers and inclusions}.
\newblock \emph{\bibinfo{journal}{J. Comput. Phys.}}
  \textbf{\bibinfo{volume}{180}}, \bibinfo{pages}{659--685}
  (\bibinfo{year}{2002}).

\bibitem{Shah2004_JBO}
\bibinfo{author}{Shah, N.}, \bibinfo{author}{Cerussi, A.~E.},
  \bibinfo{author}{Jakubowski, D.}, \bibinfo{author}{Hsiang, D.},
  \bibinfo{author}{Butler, J.} \& \bibinfo{author}{Tromberg, B.}
\newblock \bibinfo{title}{Spatial variations in optical and, physiological
  properties of healthy breast tissue}.
\newblock \emph{\bibinfo{journal}{J. Biomed. Opt.}}
  \textbf{\bibinfo{volume}{9}}, \bibinfo{pages}{534--540}
  (\bibinfo{year}{2004}).

\bibitem{Gibson2005_PhysMedBiol}
\bibinfo{author}{Gibson, A.}, \bibinfo{author}{Hebden, J.} \&
  \bibinfo{author}{Arridge, S.}
\newblock \bibinfo{title}{Recent advances in diffuse optical imaging}.
\newblock \emph{\bibinfo{journal}{Phys. Med. Biol.}}
  \textbf{\bibinfo{volume}{50}}, \bibinfo{pages}{R1--R43}
  (\bibinfo{year}{2005}).

\bibitem{Ntziachristos2005_NatBiotechnol}
\bibinfo{author}{Ntziachristos, V.}, \bibinfo{author}{Ripoll, J.},
  \bibinfo{author}{Wang, L. H.~V.} \& \bibinfo{author}{Weissleder, R.}
\newblock \bibinfo{title}{Looking and listening to light: the evolution of
  whole-body photonic imaging}.
\newblock \emph{\bibinfo{journal}{Nat. Biotechnol.}}
  \textbf{\bibinfo{volume}{23}}, \bibinfo{pages}{313--320}
  (\bibinfo{year}{2005}).

\bibitem{Svensson2007_JBO}
\bibinfo{author}{Svensson, T.}, \bibinfo{author}{Andersson-Engels, S.},
  \bibinfo{author}{Einarsd\'ott\'ir, M.} \& \bibinfo{author}{Svanberg, K.}
\newblock \bibinfo{title}{In vivo optical characterization of human prostate
  tissue using near-infrared time-resolved spectroscopy}.
\newblock \emph{\bibinfo{journal}{J. Biomed. Opt.}}
  \textbf{\bibinfo{volume}{12}}, \bibinfo{pages}{014022}
  (\bibinfo{year}{2007}).

\bibitem{Dogariu1992_WavesRandomMedia}
\bibinfo{author}{Dogariu, A.}, \bibinfo{author}{Uozumi, J.} \&
  \bibinfo{author}{Asakura, T.}
\newblock \bibinfo{title}{Enhancement of the backscattered intensity from
  fractal aggregates}.
\newblock \emph{\bibinfo{journal}{Waves Random Media}}
  \textbf{\bibinfo{volume}{2}}, \bibinfo{pages}{259--263}
  (\bibinfo{year}{1992}).

\bibitem{Sorensen2001_AerosolSciTech}
\bibinfo{author}{Sorensen, C.~M.}
\newblock \bibinfo{title}{Light scattering by fractal aggregates: A review}.
\newblock \emph{\bibinfo{journal}{Aerosol Sci. Technol.}}
  \textbf{\bibinfo{volume}{35}}, \bibinfo{pages}{648--687}
  (\bibinfo{year}{2001}).

\bibitem{Durian1991_Science}
\bibinfo{author}{{Durian}, D.~J.}, \bibinfo{author}{{Weitz}, D.~A.} \&
  \bibinfo{author}{{Pine}, D.~J.}
\newblock \bibinfo{title}{{Multiple light-scattering probes of foam structure
  and dynamics}}.
\newblock \emph{\bibinfo{journal}{Science}} \textbf{\bibinfo{volume}{252}},
  \bibinfo{pages}{686--688} (\bibinfo{year}{1991}).

\bibitem{Gittings2004_EPL}
\bibinfo{author}{Gittings, A.~S.}, \bibinfo{author}{Bandyopadhyay, R.} \&
  \bibinfo{author}{Durian, D.~J.}
\newblock \bibinfo{title}{Photon channelling in foams}.
\newblock \emph{\bibinfo{journal}{Europhys. Lett.}}
  \textbf{\bibinfo{volume}{65}}, \bibinfo{pages}{414--419}
  (\bibinfo{year}{2004}).

\bibitem{Barthelemy2008_Nature}
\bibinfo{author}{Barthelemy, P.}, \bibinfo{author}{Bertolotti, J.} \&
  \bibinfo{author}{Wiersma, D.~S.}
\newblock \bibinfo{title}{A {L}\'{e}vy flight for light}.
\newblock \emph{\bibinfo{journal}{Nature}} \textbf{\bibinfo{volume}{453}},
  \bibinfo{pages}{495--498} (\bibinfo{year}{2008}).

\bibitem{Burioni2010b_PRE}
\bibinfo{author}{Burioni, R.}, \bibinfo{author}{Caniparoli, L.} \&
  \bibinfo{author}{Vezzani, A.}
\newblock \bibinfo{title}{L\'evy walks and scaling in quenched disordered
  media}.
\newblock \emph{\bibinfo{journal}{Phys. Rev. E}} \textbf{\bibinfo{volume}{81}},
  \bibinfo{pages}{060101} (\bibinfo{year}{2010}).

\bibitem{Barthelemy2010_PRE}
\bibinfo{author}{Barthelemy, P.}, \bibinfo{author}{Bertolotti, J.},
  \bibinfo{author}{Vynck, K.}, \bibinfo{author}{Lepri, S.} \&
  \bibinfo{author}{Wiersma, D.~S.}
\newblock \bibinfo{title}{Role of quenching on superdiffusive transport in
  two-dimensional random media}.
\newblock \emph{\bibinfo{journal}{Phys. Rev. E}} \textbf{\bibinfo{volume}{82}},
  \bibinfo{pages}{011101} (\bibinfo{year}{2010}).

\bibitem{Buonsante2011_PRE}
\bibinfo{author}{Buonsante, P.}, \bibinfo{author}{Burioni, R.} \&
  \bibinfo{author}{Vezzani, A.}
\newblock \bibinfo{title}{Transport and scaling in quenched two- and
  three-dimensional {L}\'evy quasicrystals}.
\newblock \emph{\bibinfo{journal}{Phys. Rev. E}} \textbf{\bibinfo{volume}{84}},
  \bibinfo{pages}{021105} (\bibinfo{year}{2011}).

\bibitem{Burresi2012_PRL}
\bibinfo{author}{Burresi, M.}, \bibinfo{author}{Radhalakshmi, V.},
  \bibinfo{author}{Savo, R.}, \bibinfo{author}{Bertolotti, J.},
  \bibinfo{author}{Vynck, K.} \& \bibinfo{author}{Wiersma, D.~S.}
\newblock \bibinfo{title}{Weak localization of light in superdiffusive random
  systems}.
\newblock \emph{\bibinfo{journal}{Phys. Rev. Lett.}}
  \textbf{\bibinfo{volume}{108}}, \bibinfo{pages}{110604}
  (\bibinfo{year}{2012}).

\bibitem{Groth2012_PRE}
\bibinfo{author}{Groth, C.~W.}, \bibinfo{author}{Akhmerov, A.~R.} \&
  \bibinfo{author}{Beenakker, C. W.~J.}
\newblock \bibinfo{title}{Transmission probability through a {L}\'evy glass and
  comparison with a levy walks}.
\newblock \emph{\bibinfo{journal}{Phys. Rev. E}} \textbf{\bibinfo{volume}{85}},
  \bibinfo{pages}{021138} (\bibinfo{year}{2012}).

\bibitem{Lagendijk1996_PhysRep}
\bibinfo{author}{Lagendijk, A.} \& \bibinfo{author}{{van Tiggelen}, B.~A.}
\newblock \bibinfo{title}{Resonant multiple scattering of light}.
\newblock \emph{\bibinfo{journal}{Phys. Rep.}} \textbf{\bibinfo{volume}{270}},
  \bibinfo{pages}{143--215} (\bibinfo{year}{1996}).

\bibitem{Rossum1999_RevModPhys}
\bibinfo{author}{van Rossum, M. C.~W.} \& \bibinfo{author}{Nieuwenhuizen,
  T.~M.}
\newblock \bibinfo{title}{Multiple scattering of classical waves: microscopy,
  mesoscopy, and diffusion}.
\newblock \emph{\bibinfo{journal}{Rev. Mod. Phys.}}
  \textbf{\bibinfo{volume}{71}}, \bibinfo{pages}{313--371}
  (\bibinfo{year}{1999}).

\bibitem{Bouchaud1990_PhysRep}
\bibinfo{author}{Bouchaud, J.-P.} \& \bibinfo{author}{Georges, A.}
\newblock \bibinfo{title}{Anomalous diffusion in disordered media: Statistical
  mechanisms, models and physical applications}.
\newblock \emph{\bibinfo{journal}{Phys. Rep.}} \textbf{\bibinfo{volume}{195}},
  \bibinfo{pages}{127--293} (\bibinfo{year}{1990}).

\bibitem{Ben-Avraham2000}
\bibinfo{author}{Ben-Avraham, D.} \& \bibinfo{author}{Havlin, S.}
\newblock \emph{\bibinfo{title}{Diffusion and reactions in fractals and
  disordered systems}} (\bibinfo{publisher}{Cambridge University Press},
  \bibinfo{year}{2000}).

\bibitem{Olson2006_JQSRT}
\bibinfo{author}{Olson, G.~L.}, \bibinfo{author}{Miller, D.~S.},
  \bibinfo{author}{Larsen, E.~W.} \& \bibinfo{author}{Morel, J.~E.}
\newblock \bibinfo{title}{Chord length distributions in binary stochastic media
  in two and three dimensions}.
\newblock \emph{\bibinfo{journal}{J. Quant. Spectrosc. Radiat. Transfer}}
  \textbf{\bibinfo{volume}{101}}, \bibinfo{pages}{269--283}
  (\bibinfo{year}{2006}).

\bibitem{Kruijf2003_AnnNuclearrEnerg}
\bibinfo{author}{De~Kruijf, W.} \& \bibinfo{author}{Kloosterman, J.}
\newblock \bibinfo{title}{On the average chord length in reactor physics}.
\newblock \emph{\bibinfo{journal}{Ann. Nucl. Energy}}
  \textbf{\bibinfo{volume}{30}}, \bibinfo{pages}{549--553}
  (\bibinfo{year}{2003}).

\bibitem{Torquato1993_PRE}
\bibinfo{author}{Torquato, S.} \& \bibinfo{author}{Lu, B.}
\newblock \bibinfo{title}{Chord-length distribution function for two-phase
  random media}.
\newblock \emph{\bibinfo{journal}{Phys. Rev. E}} \textbf{\bibinfo{volume}{47}},
  \bibinfo{pages}{2950--2953} (\bibinfo{year}{1993}).

\bibitem{Gueron2003_PRE}
\bibinfo{author}{Gu\'eron, D.} \& \bibinfo{author}{Mazzolo, A.}
\newblock \bibinfo{title}{Properties of chord length distributions across
  ordered and disordered packing of hard disks}.
\newblock \emph{\bibinfo{journal}{Phys. Rev. E}} \textbf{\bibinfo{volume}{68}},
  \bibinfo{pages}{066117} (\bibinfo{year}{2003}).

\bibitem{Torquato2001_Book}
\bibinfo{author}{Torquato, S.}
\newblock \emph{\bibinfo{title}{Random Heterogeneous Materials: Microstructure
  and Macroscopic Properties}} (\bibinfo{publisher}{Springer},
  \bibinfo{year}{2001}).

\bibitem{Yoo1990_PRL}
\bibinfo{author}{Yoo, K.}, \bibinfo{author}{Liu, F.} \&
  \bibinfo{author}{Alfano, R.}
\newblock \bibinfo{title}{When does the diffusion-approximation fail to
  describe photon transport in random-media?}
\newblock \emph{\bibinfo{journal}{Phys. Rev. Lett.}}
  \textbf{\bibinfo{volume}{64}}, \bibinfo{pages}{2647--2650}
  (\bibinfo{year}{1990}).

\bibitem{Kop1997_PRL}
\bibinfo{author}{Kop, R. H.~J.}, \bibinfo{author}{de~Vries, P.},
  \bibinfo{author}{Sprik, R.} \& \bibinfo{author}{Lagendijk, A.}
\newblock \bibinfo{title}{Observation of anomalous transport of strongly
  multiple scattered light in thin disordered slabs}.
\newblock \emph{\bibinfo{journal}{Phys. Rev. Lett.}}
  \textbf{\bibinfo{volume}{79}}, \bibinfo{pages}{4369--4372}
  (\bibinfo{year}{1997}).

\bibitem{Elaloufi2004_JOSAA}
\bibinfo{author}{Elaloufi, R.}, \bibinfo{author}{Carminati, R.} \&
  \bibinfo{author}{Greffet, J.-J.}
\newblock \bibinfo{title}{Diffusive-to-ballistic transition in dynamic light
  transmission through thin scattering slabs: a radiative transfer approach}.
\newblock \emph{\bibinfo{journal}{J. Opt. Soc. Am. A}}
  \textbf{\bibinfo{volume}{21}}, \bibinfo{pages}{1430--1437}
  (\bibinfo{year}{2004}).

\bibitem{Bertolotti2010_AdvFunctMater}
\bibinfo{author}{Bertolotti, J.}, \bibinfo{author}{Vynck, K.},
  \bibinfo{author}{Pattelli, L.}, \bibinfo{author}{Barthelemy, P.},
  \bibinfo{author}{Lepri, S.} \& \bibinfo{author}{Wiersma, D.~S.}
\newblock \bibinfo{title}{Engineering disorder in superdiffusive {L}\'evy
  glasses}.
\newblock \emph{\bibinfo{journal}{Adv. Funct. Mater.}}
  \textbf{\bibinfo{volume}{20}}, \bibinfo{pages}{965--968}
  (\bibinfo{year}{2010}).

\bibitem{Vellekoop2005_PRE}
\bibinfo{author}{Vellekoop, I.~M.}, \bibinfo{author}{Lodahl, P.} \&
  \bibinfo{author}{Lagendijk, A.}
\newblock \bibinfo{title}{Determination of the diffusion constant using
  phase-sensitive measurements}.
\newblock \emph{\bibinfo{journal}{Phys. Rev. E}} \textbf{\bibinfo{volume}{71}},
  \bibinfo{pages}{056604} (\bibinfo{year}{2005}).

\bibitem{Svensson2010_OptLett}
\bibinfo{author}{Svensson, T.}, \bibinfo{author}{Alerstam, E.},
  \bibinfo{author}{Johansson, J.} \& \bibinfo{author}{Andersson-Engels, S.}
\newblock \bibinfo{title}{Optical porosimetry and investigations of the
  porosity experienced by light interacting with porous media}.
\newblock \emph{\bibinfo{journal}{Opt. Lett.}} \textbf{\bibinfo{volume}{35}},
  \bibinfo{pages}{1740--1742} (\bibinfo{year}{2010}).

\bibitem{Alerstam2012_PRE}
\bibinfo{author}{Alerstam, E.} \& \bibinfo{author}{Svensson, T.}
\newblock \bibinfo{title}{Observation of anisotropic diffusion of light in
  compacted granular porous materials}.
\newblock \emph{\bibinfo{journal}{Phys. Rev. E}} \textbf{\bibinfo{volume}{85}},
  \bibinfo{pages}{040301} (\bibinfo{year}{2012}).

\bibitem{Wang1995_CMPB}
\bibinfo{author}{Wang, L.}, \bibinfo{author}{Jacques, S.} \&
  \bibinfo{author}{Zheng, L.}
\newblock \bibinfo{title}{{MCML} {M}onte {C}arlo modeling of light transport in
  multilayered tissues}.
\newblock \emph{\bibinfo{journal}{Comput. Meth. Prog. Bio.}}
  \textbf{\bibinfo{volume}{47}}, \bibinfo{pages}{131--146}
  (\bibinfo{year}{1995}).

\bibitem{Sadjadi2011_PRE}
\bibinfo{author}{Sadjadi, Z.} \& \bibinfo{author}{Miri, M.}
\newblock \bibinfo{title}{Diffusive transport of light in two-dimensional
  granular materials}.
\newblock \emph{\bibinfo{journal}{Phys. Rev. E}} \textbf{\bibinfo{volume}{84}},
  \bibinfo{pages}{051305} (\bibinfo{year}{2011}).

\end{thebibliography}
\end{document}